\documentclass{optica-article}

\journal{opticajournal} 

\articletype{Research Article}

\usepackage{lineno}

\begin{document}

\title{Single Plane Spatial Mode Sorter}

\author{Khen Cohen,\authormark{1,*} Yoav Yosif-Or,\authormark{2}, Yaron Oz, \authormark{1} and Ady Arie\authormark{2}}

\address{\authormark{1}School of Physics and Astronomy, Tel-Aviv University, Tel-Aviv 69978, Israel\\
\authormark{2}School of Electrical and Computer Engineering, Tel-Aviv University, Tel-Aviv 69978, Israel}

\email{\authormark{*}khencohen@mail.tau.ac.il} 


\begin{abstract*} 
A mode sorter separates a set of $M$ orthogonal spatial modes in a shared input channel into $M$ different output channels. Here we present an analytic derivation and experimental validation of a single plane device for sorting spatial modes from a diverse variety of mode families, including Hermite-Gaussian (HG), Laguerre-Gaussian (LG), Bessel-Gaussian (BG), with almost no cross-talk. This sorting capability is required for a wide range of applications that employ classical or quantum light.
We also show that applying this design in order to sort {\color{black}a} set of Orbital Angular Momentum (OAM) modes with zero radial index reproduces the well-known Fork grating configuration.
Furthermore, by taking the limit of $M \rightarrow \infty$, we present an analytical expression for sorting all the modes of a given family.
By operating this device in reverse, it can be used to generate arbitrary modes, by illuminating it with a Gaussian beam.
The power transmission coefficient for this sorter goes as $O(M^{-1})$ and we provide a mathematical proof that this is optimal {\color{black} for any typical arrangement of the detector positions}. We further study the sorter sensitivity to wavelength and random phase noise.
\end{abstract*}


\section{Introduction}
\label{sec:Introduction}

Recent advances in optical technologies have driven a growing need for high-dimensional encoding in optical systems \cite{erhard2020advances}, particularly for applications such as spatial-division multiplexing in classical optical communications \cite{yan2014high, bozinovic2013terabit}, super-resolution imaging \cite{hell1994breaking, tsang2016quantum}, quantum computing, and quantum key distribution (QKD) \cite{scarani2009security}. Spatial modes \cite{brandt2020high}, eigenstates of propagation media such as free space or optical fibers, represent a versatile approach for encoding information across multiple dimensions \cite{Otte20}. These modes, including Laguerre–Gaussian (LG), Hermite–Gaussian (HG), and Bessel–Gaussian (BG) patterns, have been widely adopted across diverse fields: electron beam manipulation \cite{GarcElectronBeams}, free-space and fiber-based optical communication \cite{chan2006free, zhou2021high}, continuous-variable quantum computing \cite{fabre}, resonant-cavity mode control \cite{Cavity}, and more.

The ability to sort spatial modes is critical across both classical and quantum domains. In spatial-division multiplexing, efficient mode discrimination reduces inter-channel crosstalk and enables terabit-scale data transmission \cite{bozinovic2013terabit}. In super-resolution imaging, tailored projection onto spatial basis functions allows one to overcome the diffraction limit and extract sub-diffraction features \cite{hell1994breaking, tsang2016quantum}. In quantum communications, mode sorting facilitates direct detection of high-dimensional QKD alphabets without the need for full quantum tomography, thereby improving key rates and reducing system complexity \cite{LvovskyCVTomography}. Furthermore, in quantum state characterization, spatial-mode sorters can replace exhaustive projective measurements with targeted modal projections, substantially reducing the measurement overhead required for tomography of entangled spatial states \cite{mair2001entanglement}.

Despite the considerable advantages of spatial mode encoding, several technical challenges exist in manipulating these modes within optical systems. Primary among these are sorting the modes \cite{Alvaro23, Defienne20, Kupianskyi23, Alarcon23}, spatial mode conversion \cite{ModeConveter}, and mode generation \cite{ModeGeneration1, ForbesModeGeneration2}. These challenges increase dramatically when dealing with large numbers of modes or arbitrary mode types.

The evaluation of optical mode sorters typically relies on three fundamental metrics: complexity (in terms of optimization or calibration requirements), power transmission factor (describing the ratio between sorted and input intensity), and cross-talk (defined as the ratio between correctly sorted intensity and total transmitted intensity). Current research indicates that effectively sorting M spatial modes using traditional approaches typically requires approximately 2M+1 optical layers \cite{Morizur2010}. While several studies have demonstrated high-performance spatial mode sorting with minimal cross-talk and high power transmission, these implementations often require multiple cascaded optical elements, ranging from 4 to 7 layers or possibly more \cite{fontaine2019laguerre}.

One important method for sorting modes is the fork grating \cite{Fork}, which is a key method to sort Optical Angular Momentum (OAM) beams, with theoretically zero cross talk. However, this method is limited to modes with zero radial index \cite{bazhenov1990laser}. The Fork Grating was also shown in the quantum regime \cite{mair2001entanglement}, and also was used to sort electron beams \cite{mcmorran2012platelet, verbeeck2012new}. Alternative approaches include nonlinear optics \cite{bloch2012twisting}, or the Berkhout log-polar transformation \cite{Berkhout}, which requires only two layers but exhibits lower cross-talk performance and is limited to {\color{black} OAM} modes. Single-layer solutions such as the Angular Lens \cite{AngularLens} have been employed to sort OAM modes, but their performance remains limited to radial index of zero, and with cross-talk typically exceeding acceptable thresholds even for small mode counts. 

While existing methods achieve high performance either through multiple optical layers or by restricting themselves to specific spatial mode types, designing a single-layer sorter capable of sorting arbitrary mode sets with high quality remains an open challenge in the field.

In this work, we present a novel single-layer sorter that achieves high-performance sorting with nearly zero cross-talk for sets of $M$ orthonormal spatial modes. This performance comes with an arbitrary trade-off in intensity transmission coefficient, which typically scales as $O(M^{-1})$. We extend our method to accommodate non-orthogonal mode sets and provide a comprehensive analysis of its performance characteristics. Our findings are supported by both numerical simulations and experimental validation across three distinct spatial mode types: Laguerre-Gaussian, Hermite-Gaussian, and Bessel-Gaussian. Furthermore, by taking the limit of {\color{black}an} infinite number of modes $M \rightarrow \infty$ we show a closed analytical form of a single sorter to sort all the modes in a group family at once. The proposed method offers an intuitive approach that is straightforward to design and implement experimentally, representing an effective compromise for various experimental setups.

The remainder of this paper is organized as follows: Section \ref{sec:Methods} introduces the theoretical model, describes our proposed sorter, and details the experimental setup. Section \ref{sec:Results} presents our results, evaluating different spatial mode sets \ref{sub:ExperimentalValudation}, including Mutually Unbiased Basis (MUB) configurations \cite{MUB}, and provides noise and wavelength analysis \ref{sub:WavelengthNoiseAnalysis}, followed by proposed spectroscopic application. We demonstrate that the Fork sorter for OAM modes represents a specific case of our more general sorting approach. An application of a mode generator is also presented in this section.
In section \ref{sec:infinity_modes}, we discuss and deal with the case of taking {\color{black}an} infinite number of modes.
Section \ref{sec:Discussion} discusses our findings, addresses the limitations and the potential of our method. Finally, Section \ref{sec:Conclusion} concludes the paper and suggests directions for future research, with additional technical details provided in the Supplementary Materials.

\begin{figure*}
    \centering
    \includegraphics[width=1.0\linewidth]{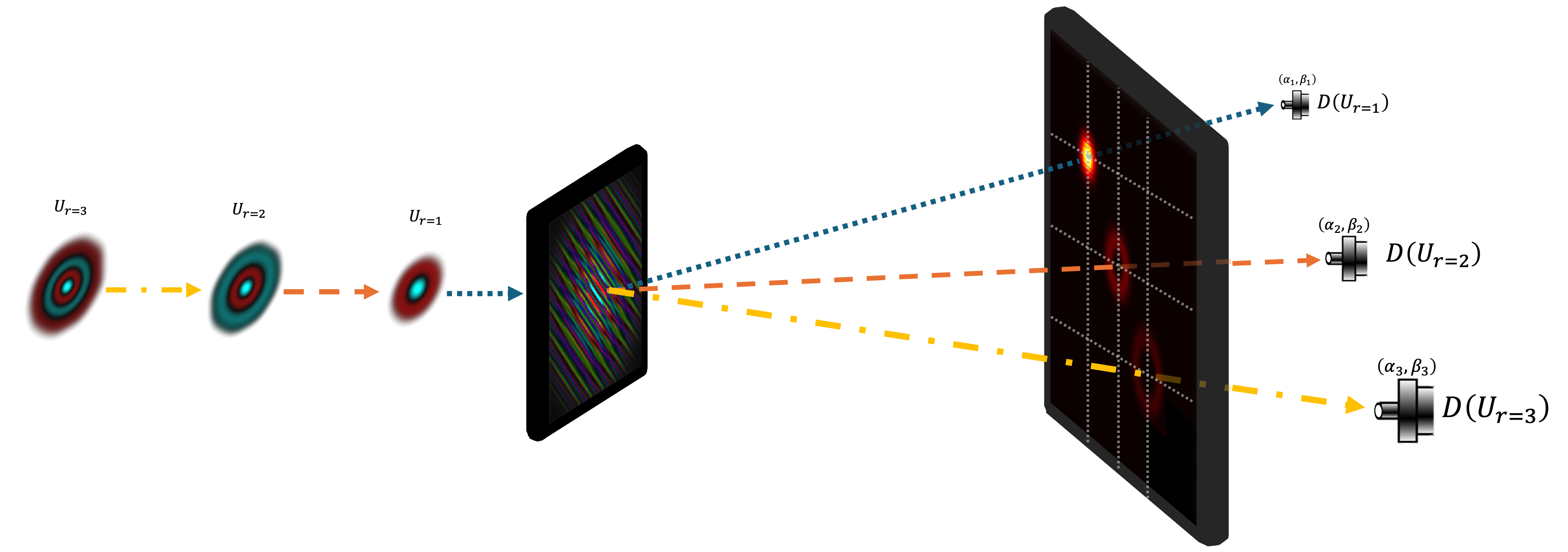}
    \caption{A schematic diagram of the setup. Distinct spatial modes traverse a single plane of a Spatial Light Modulator (SLM). Accompanied by scattering, each spatial mode transmits its intensity directly to its corresponding detector, whereas the other detectors register zero intensity. For instance, the mode $U_{r=1}$ (depicted as dashed {\color{black} blue line}) passes through the SLM and is registered by detector $D\left(U_{r=1}\right)$, while the other detectors detect zero intensity. In each mode, scattering occurs around the detectors of other modes with a characteristic size of the bandwidth $\Omega$, and the light is filtered out prior to reaching the detector.}
    \label{fig:ExpSetup}
\end{figure*}

\section{Methods}
\label{sec:Methods}
In this section, we present the theoretical framework and the suggestion of the sorter. An extended derivation is shown in the Supplementary Material. Second, we present the experimental setup we used in order to examine our sorter performance.

\subsection{Theoretical Model}
\label{sub:TheoreticalModel}

We consider the case of sorting a set of $M$ spatial tokens, such as spatial modes e.g. Hermite Gauss, Laguerre–Gaussian or Bessel Gauss set of modes, which we denote as $\{f_m(x,y)\}_{m=1}^M$, with a single sorter mask $S(x,y)$. The output of the system, will be considered via taking the far field approximation of their multiplication, according to \cite{goodman2005introduction}:
\begin{gather}
    E_m(\xi, \eta) = \int \int f_m(x,y) S(x,y) e^{-i\frac{2\pi}{\lambda z}\left( \xi x + \eta y\right) } dx dy \ ,
\end{gather}
while the coordinates $\xi, \eta$ are two spatial coordinates in the far field, and output intensity is given by $I_m(\xi, \eta) = |E_m(\xi, \eta)|^2$ .

In the output plane, we consider $M$ spatially separated detectors in locations $\left\{(\alpha_\mu, \beta_\mu)\right\}_{\mu=1}^M$, each is related to another input mode.

We define the captured field amplitude at each detector $\mu$ as:
\begin{gather}
    E_{m,\mu}= \int \int f_m(x,y) S(x,y) e^{-i\frac{2\pi}{\lambda z}\left( \alpha_\mu x + \beta_\mu y\right) } dx dy \ .
\end{gather}
A sorter $S(x,y)$ would be considered as a good sorter if the tensor $E_{\mu,m}$ will be as close to diagonal as possible, namely, maximum for $\mu=m$ and zero for $\mu \neq m$.

Working in the modes spatial basis, our proposed sorter is defined as follows:
\begin{gather}
\label{eq:sorter}
    S(x,y) = \frac{1}{\sqrt{M}} \sum_{m^\prime=1}^M f_{m^\prime}^*(x,y) e^{i\frac{2\pi}{\lambda z} \left( \alpha_{m^{\prime}}x +\beta_{m^\prime}y \right)} \ .
\end{gather}
We refer the reader to an extended theoretical derivation in the Supplementary Material. In the typical case of a sorting set of orthogonal modes, the sorting crosstalk is approximately $0\%$, but with a price of power transmission which goes as $~O(M^{-1})$.
{\color{black} The Supplementary Material presents a mathematical proof showing that, for any typical arrangement of detector positions (averaged over all possible locations), our sorter attains the optimal performance for a single layer. Put differently, once the power detector positions are randomly chosen and fixed, no further enhancement of power transmission or reduction of cross-talk is possible.
Numerical results indicate that the overall power transmission coefficient concentrates sharply around its mean value, exhibiting a very small variance (about $10^{-7}$), and therefore effectively behaves as a deterministic quantity. Moreover, the power transmission is bounded from below by $M^{-1}$, and can only be marginally improved using optimization over the detectors locations.}

For non-orthogonal set of modes, the cross-talk is proportional to the average inner product between any two modes in the set:
\begin{gather}
    CT = \frac{1}{M} \sum_{\mu > m} \left|\langle {f_\mu} | {f_m}\rangle \right|^2 \ .
\end{gather}
Our method gives a clear and intuitive way to get $I_{\mu, m} \approx \frac{1}{M}\delta_{\mu, m}$.

The separation between two detectors need to be larger than the maximum bandwidth over all the modes $\Omega$. Where for Gaussian beam the bandwidth can be evaluated as:  $\Omega\approx \frac{\lambda z}{2w \bar{M} }$, while $w$ is the width of the incidence beam and $\bar{M}$ is the minimum $M$ factor over all the modes.

Therefore, the minimum distance between any two detectors need to be:
\begin{gather}
    \forall \mu, m \ , \ \sqrt{ (\alpha_\mu - \alpha_m)^2+(\beta_\mu - \beta_m)^2   }\geq \Omega \ .
\end{gather}

\begin{figure*}
    \centering
    \includegraphics[width=1.0\linewidth]{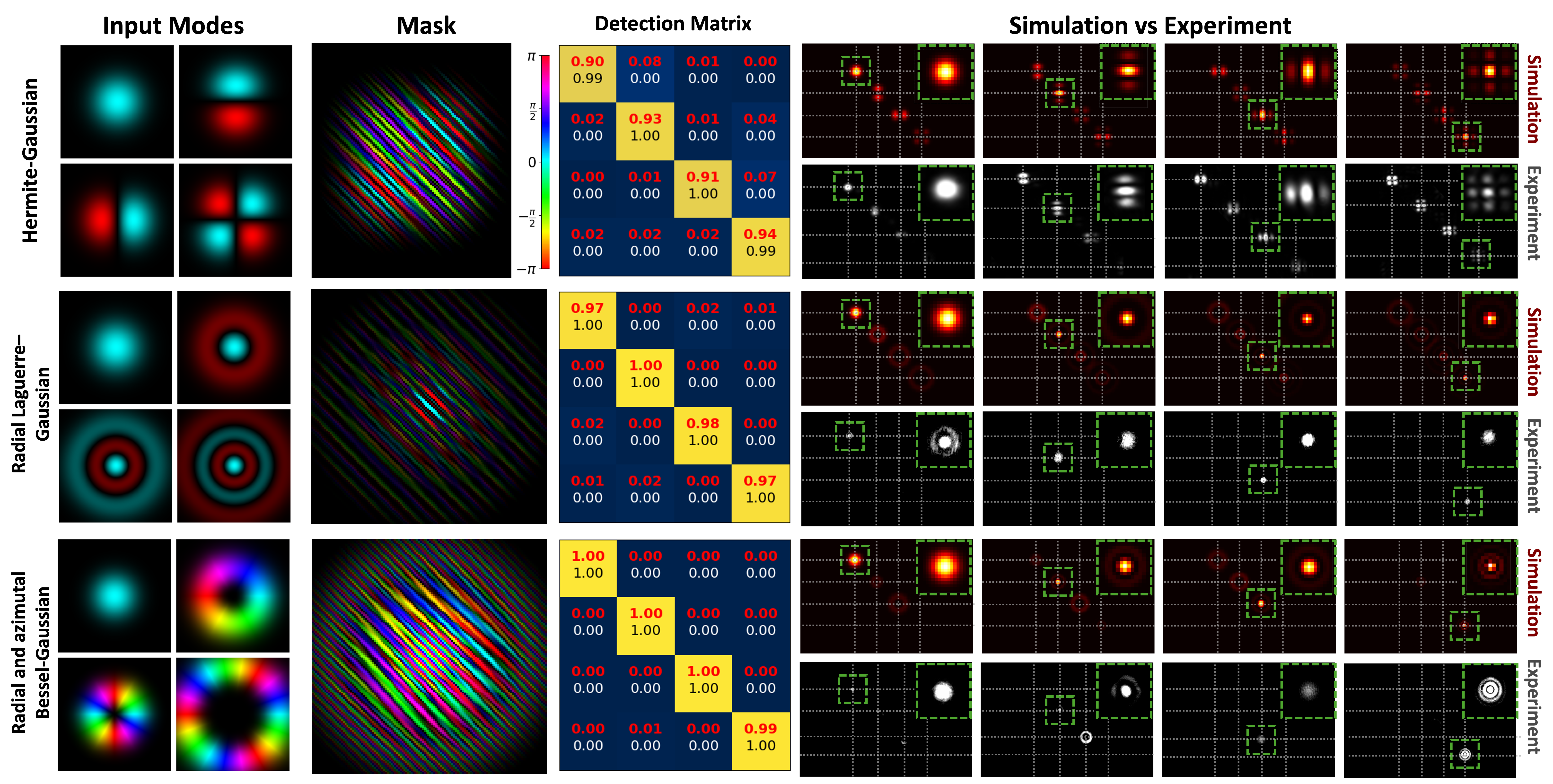}
    \caption{Sorting results for three {\color{black}groups} of modes. From top to bottom: (a) Hermite-Gaussian modes $HG_{0,0},HG_{1,0},HG_{0,1}, HG_{1,1}$ ; (b) Radial Laguerre–Gaussian modes $LG_{0,0},LG_{1,0},LG_{2,0},LG_{3,0}$; and (c) Bessel-Gaussian modes $BG_{0,0},BG_{1,1},BG_{2,-2},BG_{3,3}$. The modes and the sorter size are about three times the beam waist. The detection matrices report the experimental result in red, and simulation results in white. The values in the detection matrix were determined by evaluating a small vicinity around the detector’s expected location - typically within one to three pixels.}
    \label{fig:mainResults}
\end{figure*}
\subsection{Experimental Implementation}
\label{sub:ExperimentalSetup}

As the light source, we used a $632.8$ nm, $2$ mW, unpolarized helium-neon laser, followed by a linear polarizer. The beam was expanded using a beam expander consisting of two plano-convex lenses with focal lengths of $50$ mm and $500$ mm, separated by $550$ mm, resulting in a $10\times$ magnified and collimated output. This expanded beam illuminated a phase-only spatial light modulator (SLM-A), with the beam diameter adjusted to cover over $90\%$ of the SLM’s active area to maximize spatial resolution. SLM-A was programmed via computer to encode the desired spatial mode, utilizing the Bolduc method for phase-only encoding \cite{Bolduc13}. 

After reflection from SLM-A, the beam propagated through a $4f$ system formed by two $500$ mm focal length lenses. At the Fourier plane of this system, an iris aperture was placed to spatially filter out all but the first diffraction order, thus isolating the encoded mode. A second SLM-B, acts as the sorter, is located then, followed by another $2f$ system, to get the far field projected into the camera. Please note that our implementation uses a phase-only sorter, despite the theoretical requirement for a sorter involving both phase and amplitude. The rationale behind this decision is examined in the Supplementary Material, along with a discussion of the differences, in the discussion section.
A schematic diagram of the setup is shown in Figure \ref{fig:ExpSetup}.

\section{Results}
\label{sec:Results}
Here, we present both simulation and experimental results obtained using the proposed sorter. We evaluated its performance across several types of spatial modes, including a four-mode sorting configuration. Our findings indicate that the sorter performs well even for a larger number of modes; however, this comes at the cost of power loss. We demonstrate that our sorter extends the well-known Fork sorter to accommodate different OAM-generated modes. Attempts to sort mutually unbiased bases (MUBs) yielded a cross-talk, as expected due to the lack of orthogonality among the modes. We analyze the impact of random noise and wavelength variations on the sorter's performance, followed by a proposed spectroscopy technique. We further show a closed analytical form of sorter that sort all modes at once, and we show a mode-generator application by inverting the direction of light.

\subsection{Experimental Validation}
\label{sub:ExperimentalValudation}

We tested the sorter on four spatial modes from several mode families: Laguerre-Gaussian modes with varying radial indices, Hermite-Gaussian modes, Bessel-Gaussian modes, and mixed modes combining Laguerre-Gaussian and Hermite-Gaussian profiles. The results are presented in Figure \ref{fig:mainResults}. We further evaluated its performance with different OAM-generated modes, as described in Section \ref{sub:OAM}, and with mutually unbiased bases (MUBs), as shown in Figure \ref{fig:mub}. The mathematical definitions and key properties of these mode families are provided in the Supplementary Material.

The sorter was designed to spatially separate the output modes by approximately 1 mm, arranged along a diagonal line at the output plane. However, this configuration of the detectors, is easily adjustable using $\alpha_m$, and $\beta_m$ coefficients as described in Eq. \ref{eq:sorter}. In each experiment, the sorter pattern was displayed on SLM-B, while the input mode was varied using SLM-A. The resulting intensity distribution, captured by the camera, exhibited a distinct peak at the detector's location corresponding to the sorted mode, along with some surrounding intensity pattern. Since {\color{black}we} consider only detection at the designated points, we can ignore the surrounding pattern. The transmission matrix was computed under the assumption of a $1 \times 1$ pixel detector size and normalized by the total transmitted power.

The normalized transmission matrix should ideally be a unity matrix, hence the diagonal elements represent the sorting efficiency of the desired mode $U_n$ to the designated detection point, ($\alpha_b$, $\beta_n$) whereas the off-diagonal elements represent the cross-talk. The average value of the efficiency for the three mode families, each with 4 modes, is $96.6\%$, and the average cross talk is $2.7\%$.

\subsection{Sorting modes with different Orbital Angular Momentum}
\label{sub:OAM}
{\color{black}We find that} designing the sorter for different {\color{black} OAM} modes yields the familiar fork-like structure \cite{Fork, bloch2012twisting, verbeeck2012new, mcmorran2012platelet}. This observation reinforces the understanding that our proposed sorter generalizes the widely adopted fork sorter.
In particular, similar performance is observed for other OAM-generated modes as well. An illustrative example and corresponding experimental results are presented in the Supplementary Materials.

\subsection{Power transmission analysis}
\label{sub:PowerTransmission}
\begin{figure}
    \centering
    \includegraphics[width=0.75\linewidth]{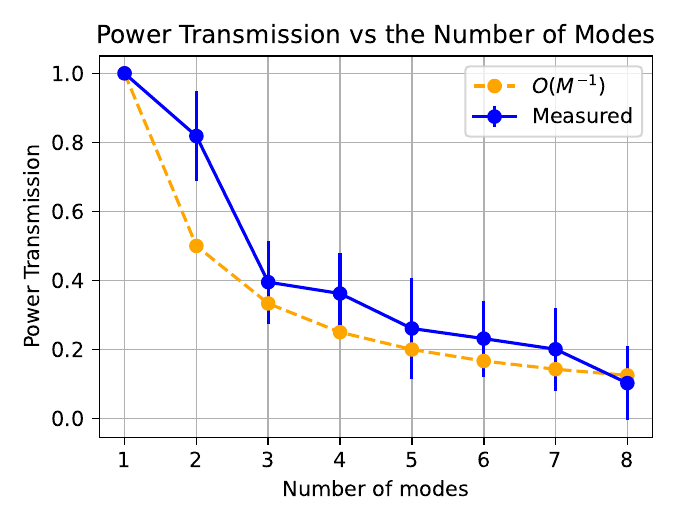}
    \caption{Experimental results of the relationship between the power transmission coefficient and the number of modes. Each data point signifies an average over all sorted modes. The standard deviation was computed by combining two independent sources of statistical noise: $\sigma = \sqrt{ \sigma_{\text{stat}}^2 + \sigma_{\text{meas}}^2 }$.}
    \label{fig:PowerTransmission}
\end{figure}

According to the theoretical derivation (see the Supplementary Material), the anticipated power transmission coefficient declines in a manner inversely proportional to the number of modes, expressed as $P_{T} \sim O(M^{-1})$, which is optimal {\color{black} as discussed in \ref{sub:TheoreticalModel}}. Figure \ref{fig:PowerTransmission} presents an experimental comparison of the observed power transmission coefficient. The assessment of the transmission coefficient was conducted by comparing the ratio of the peak values of the sorted mode (aligned at its position) to the peak value of unsorted power (where the SLM acts as a mirror). Every value depicted in the graph was determined as an average across all sorted modes. The standard deviation for each power transmission was derived by calculating the standard deviation of its power transmission coefficient, denoted as $\sigma_{\text{stat}}$, with an additional independent noise source associated with the measuring device, which we approximated as $\sigma_{\text{meas}} \approx 0.1$, {\color{black} estimated from repeated measurements and showing no significant dependence on the measured value}.

\subsection{Wavelength and Phase Noise Robustness}
\label{sub:WavelengthNoiseAnalysis}
We evaluated the performance of the mode sorter under two primary noise sources: wavelength deviation and Gaussian phase noise introduced within the sorter. Both phase noise and wavelength analysis are presented in details in the Supplementary Material.

Notably, the sorter demonstrates robustness under moderate noise levels, particularly in the case of Hermite-Gaussian mode sorting, provided the noise standard deviation remains below approximately $0.5\pi$. While some modes appear more sensitive than others, it is seen that small noise levels (e.g., $\sigma \sim 0.1\pi$) do not significantly degrade performance for any of them. We attribute this robustness to the uncorrelated nature of the phase noise, which distributes incoherently across the output plane -disrupting the field locally but not coherently redirecting energy toward incorrect detector positions.  

While multi-layer methods are highly sensitive to wavelength variations due to the accumulation of errors across different layers, our single-layer design exhibits only a scaling effect on the projected pattern, which can be easily corrected.

Wavelength sensitivity analysis indicates that the sorter is generally robust to small variations in wavelength, particularly near the optical axis (for small $\alpha$ and $\beta$ in the theoretical analysis). However, it is important to note that for large sorting angles, typically arising when a large number of spatial modes $M$ are employed, the sensitivity to wavelength increases. This occurs because larger sorting angles produce more spatially localized intensity patterns at the detector plane, making them more susceptible to shifts induced by wavelength variations.

This inherent wavelength sensitivity for large angles can be further exploited to enable spectroscopic functionality. Specifically, for a sorter designed to operate at a base wavelength $\lambda_0$, with a nominal sorting distance of $a_0$ from the optical axis, the difference in the wavelength is given by:
\begin{equation} \label{eq:spectroscopy}
    \Delta \lambda  = \lambda_0 \left( \frac{1}{1+\tfrac{d}{a_0} }-1 \right) \ ,
\end{equation}
Note that both $\Delta \lambda$ and $d$ may take positive or negative values.

To assess our proposed spectroscopy technique, we utilized the sorter for Bessel-Gaussian modes. Each time, the sorter was fine-tuned for another individual wavelengths and consistently employed a 633$_{nm}$ laser for each experiment. For each sorting configuration, we determined the distance from the optical axis, and calculated the presumed $\Delta \hat{\lambda}$ (using eq. \ref{eq:spectroscopy}) from this data. The outcomes are presented in table \ref{tab:spectroscopyShort}.
We kindly refer the reader to the Supplementary Material for a complete derivation, and the experimental details.
\begin{table}
    \centering
    \begin{tabular}{|c|c|c|c|c|l|}\hline
 $\Delta \lambda _{ \text{nm} }$& 0& 7& 17& 27& 37\\\hline
 $\Delta\hat{ \lambda }_{ \text{nm} }$& 0& 5.577& 17.604& 27.631& 37.424\\ \hline
    \end{tabular}
    \caption{Spectroscopy results for sorting Bessel-Gaussian modes ($BG_{0,0}$, $BG_{1,1}$, $BG_{2,-2}$, and $BG_{3,3}$), with a focus on the mode $BG_{1,1}$. $\Delta \hat{\lambda}$ represents the estimated wavelength shift inferred from the pixel displacement.}
    \label{tab:spectroscopyShort}
\end{table}

\subsection{Mutually unbiased bases}
\label{sub:MUB}
\begin{figure}
    \centering
    \includegraphics[width=0.7\linewidth]{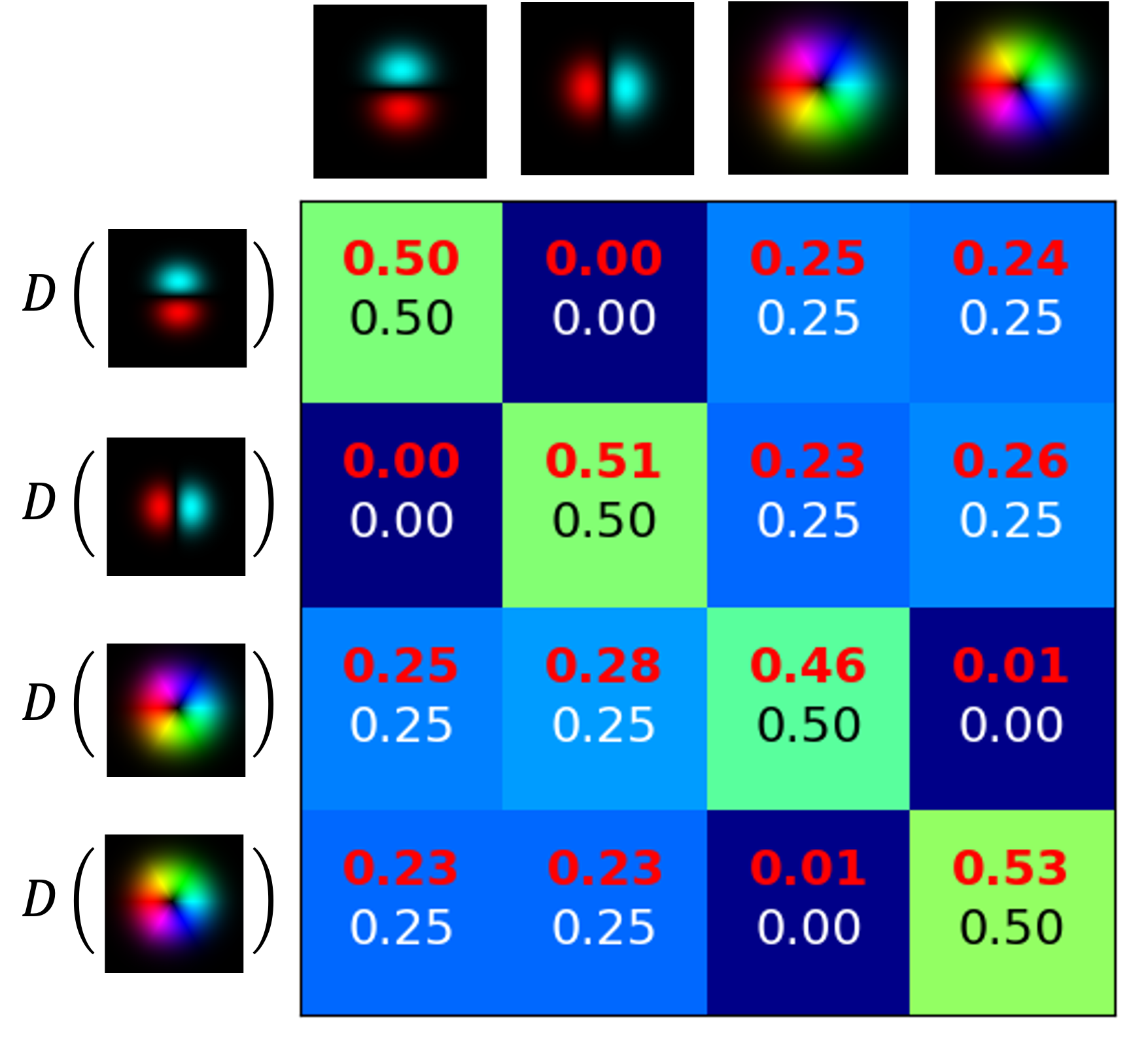}
    \caption{MUB confusion matrix. The function \( D(.) \) denotes the detector linked to each transmitted mode. In each cell, the value at the top (red) indicates the experimental data, while the value at the bottom (white/black) reflects the simulation outcome. Refer to the Supplementary Materials for the projection figures.}
    \label{fig:mub}
\end{figure}

As discussed in the theoretical analysis (see Section \ref{sub:TheoreticalModel}), the cross-talk performance of the sorter depends on the inner products between the modes in the set. This introduces a limitation to our model, particularly when attempting to sort mutually unbiased bases (MUBs), which are especially relevant for quantum information processing applications. To investigate this, we considered a set of four MUB spatial modes constructed from HG and LG modes $\left\{ HG_{1,0}, HG_{0,1}, LG_{0,1}, LG_{0,-1} \right\}$, while it holds that:
\begin{gather}
    LG_{0,\pm 1} = \frac{1}{\sqrt{2}}\left(HG_{1,0}\pm i HG_{0,1} \right) \ .
\end{gather}
Figure \ref{fig:mub} shows that the sorter's performance for each detector (rows), and injected mode (cols), for the experimental and simulation results. An alternative design (not shown) is to have a sorter for all the 4 modes in the mutually unbiased basis of the Hermite-Gauss modes. In that case, each of these modes at the input would be detected in single output detection point, whereas the energy of input Hermite-Gauss modes would be distributed among multiple output points. In the Supplementary Materials we provide additional details on the experimental result.

\subsection{Mode Generation}
\label{sub:mode_generation}
\begin{figure}[t]
    \centering
    \includegraphics[width=0.7\linewidth]{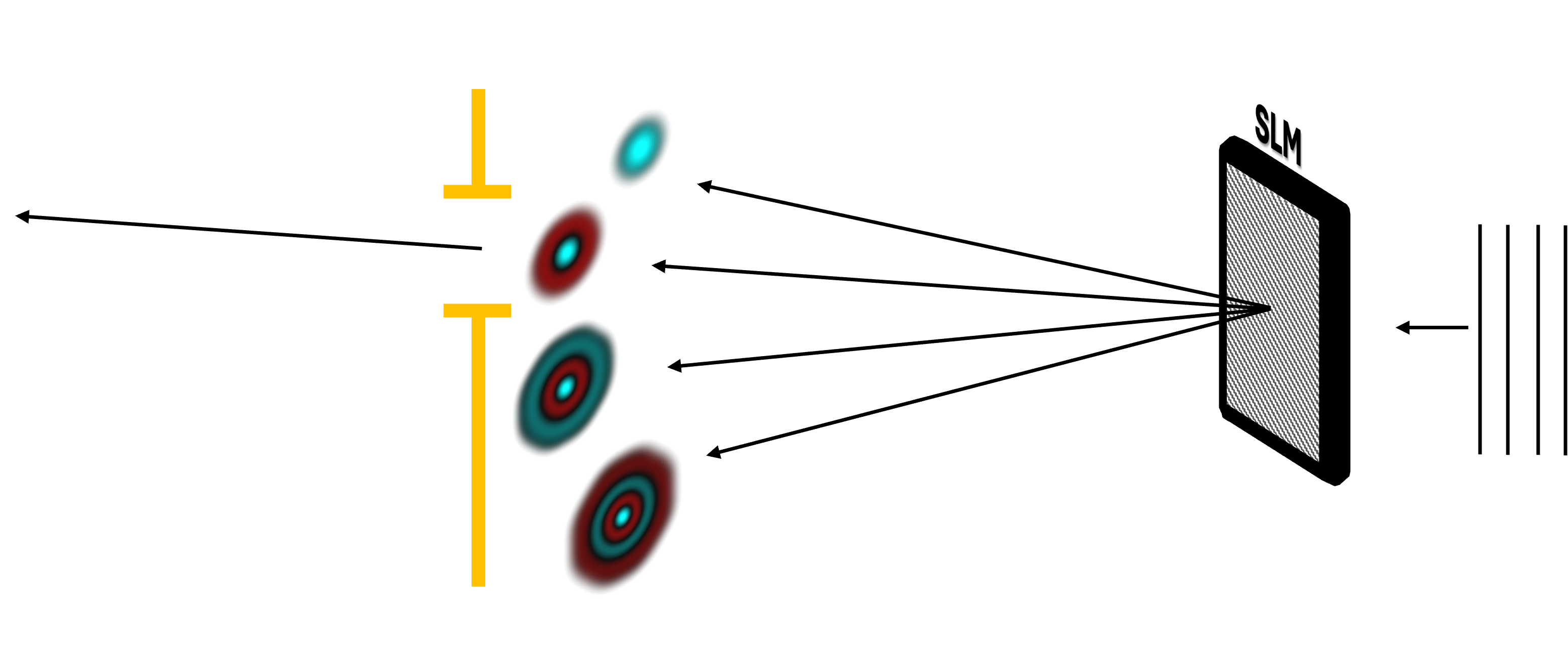}
    \includegraphics[width=0.7\linewidth]{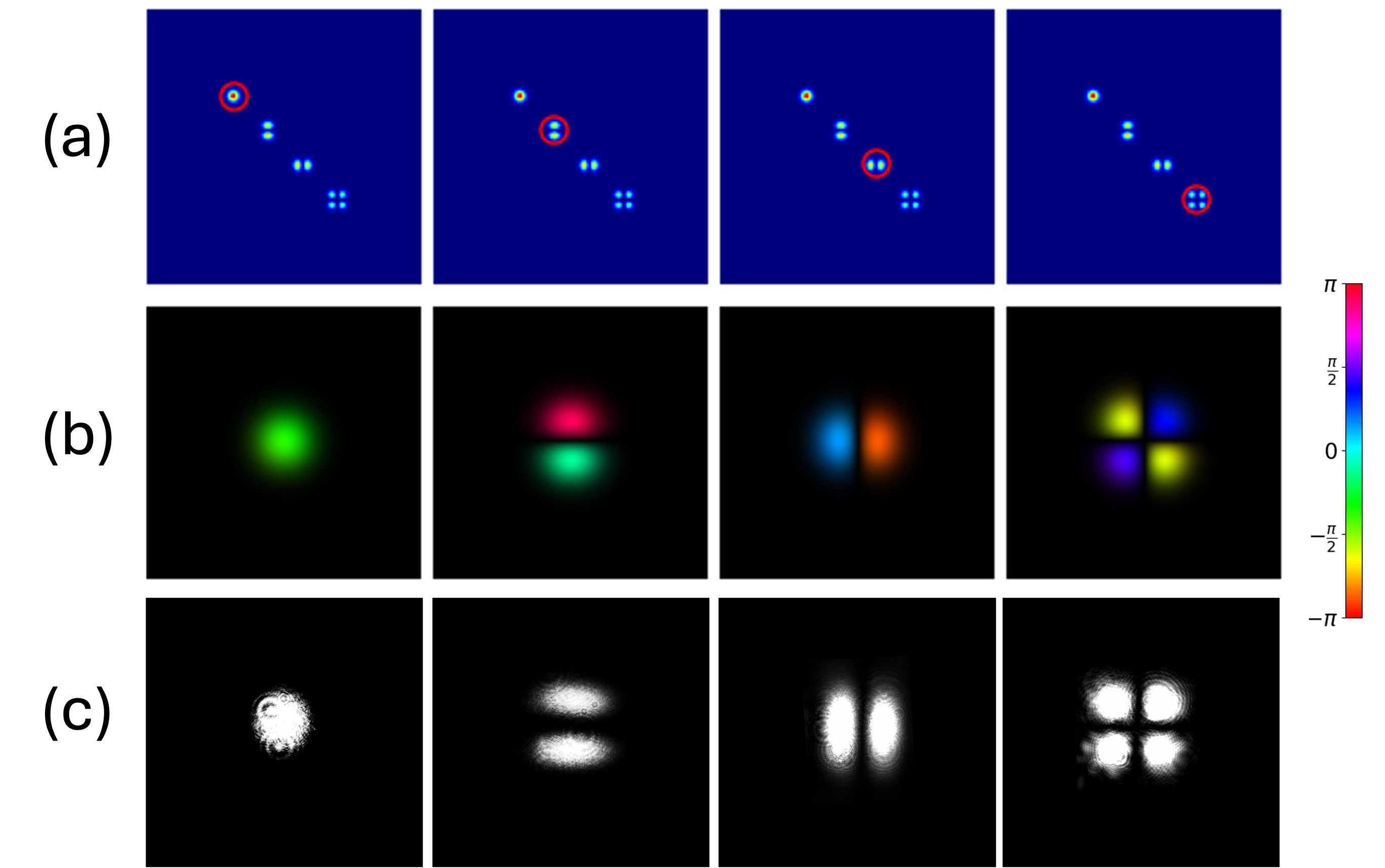}
    \caption{Using the sorter in reverse for spatial mode generation. Top: Schematic of the setup, where a plane wave is incident on the sorter in the reverse direction, followed by a spatial filter and a subsequent far-field propagation stage. Bottom: Illustration of mode generation for four spatial modes: $HG_{0,0}$, $HG_{0,1}$, $HG_{1,0}$, and $HG_{1,1}$. (a) Far-field intensity pattern produced by the mask; the red circles denote the spatial filter selecting each mode. (b) The target (true) spatial modes. (c) The corresponding generated modes. While phase information is not visually apparent in the generated intensity patterns, it was correctly reconstructed and validated using inner product measurements.}
    \label{fig:modeGenerator}
\end{figure}

Another direct application of our sorter is spatial mode generation. By projecting a plane wave, or a broad Gaussian beam, onto the mask, the far-field pattern yields the Fourier transforms of the target modes, each emerging at a distinct angle. By spatially filtering these components and applying a subsequent Fourier transform, the desired mode can be reconstructed. Thus, as demonstrated in Figure \ref{fig:modeGenerator}, the sorter can conveniently function as a mode generator.

\section{Extension to all modes}
\label{sec:infinity_modes}

Extending the sorter as presented in equation \ref{eq:sorter} to infinite number of modes $M \rightarrow \infty$, and define the spacing between the detectors as $\alpha_m = a m$, and $\beta_m = b m$ :
\begin{gather}
    S_{M \rightarrow \infty}(x,y) = \sum_{k=1}^{\infty} f_{k}^*(x,y) \  t ^{k} \ ,
\end{gather}
While we define $ t\equiv e^{i\frac{2\pi}{\lambda z} \left( ax +b y\right)}$.
This expression corresponds to a generating function and can therefore be evaluated analytically for specific choices of mode sets. The full derivation is provided in the Supplementary Material. Here, we present closed-form solutions for several specific cases: Hermite-Gaussian (HG) modes, Laguerre-Gaussian (LG) modes with fixed radial number, {\color{black} OAM} modes, and Bessel-Gaussian (BG) modes.

\begin{figure*}
    \centering
    \includegraphics[width=1.0\linewidth]{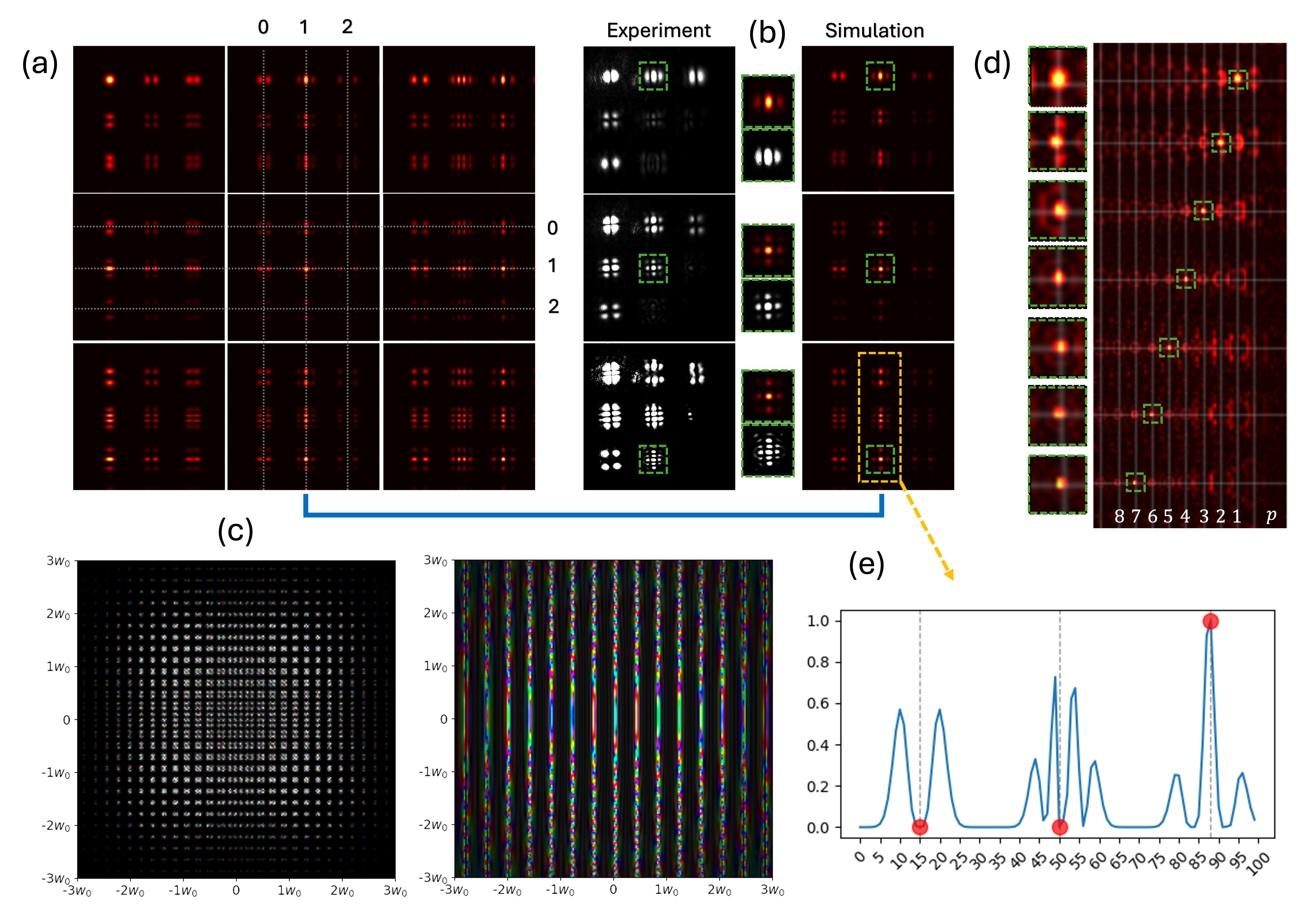}
    \caption{The analytical mode sorter used for distinguishing between spatial modes of Hermite–Gaussian and radial Laguerre–Gaussian beams. (a) HG mode sorting projects any $(n,m)$ mode to a grid location $(a n, b m)$. (b) Comparison between simulation and experimental results for sorting $HG_{1,0}$, $HG_{1,1}$, and $HG_{1,2}$. (c) Sorter masks: HG (left) and radial LG (right). (d) Radial LG sorting projects modes with radial index $p$ to positions $a p$. (e) Power transmission is high at the corresponding detector and approaches zero at the others. Notably, the performance depends on resolution: higher resolution yields a more localized (delta-like) response at the correct detector position.}
    \label{fig:analyticalSorter}
\end{figure*}

The analytical expression for each of the sorters is shown in Table \ref{tab:infinity_sorters}. The results are shown in Figure \ref{fig:analyticalSorter}, the result for HG in (c), the results for radial LG in (b), and the masks in (a).

It is important to mention that as the number of modes approaches infinity, the power transmission coefficient diminishes to zero, corresponding to the $O(M^{-1})$ scaling. This scenario is not observed in reality because it demands infinite frequency modulation by the sorter. In practice (as shown in the simulation), the physical device implementing the sorter has a maximum frequency constrained by the optics or pixel resolution. This restriction limits the support of higher order sorting, and realistically, such modes are not regarded as sorted modes. Therefore, one can infer that an increase in the spatial resolution supported by the sorter results in a greater number of effectively sorted modes, and the power transmission decays only according to these supported modes.

\begin{table*}
    \centering
    \begin{tabular}{|c|c|c|} \hline 
         Type&  Sorter&  Detectors $(x,y)$\\ \hline 
         $HG_{n,m}$&  $
            S_{\mathrm{HG}}(x,y) = \sqrt{\frac{2}{\pi}} \, \frac{1}{w_0} \, e^{  -\frac{x^2 + y^2}{w_0^2} } e^{  \frac{2}{w_0} (x e^{-iax} + y e^{-iby}) - e^{-2iax} - e^{-2iby} } \nonumber
            $
&   $(an,bm)$\\ \hline 
         $LG_p$ radial&  $
S_{\mathrm{LG}}(x,y) = \sqrt{\frac{2}{\pi}} \, \frac{1}{w_0 (1-e^{iax-\varepsilon x})} e^{ -\frac{x^2+y^2}{w_0^2} \frac{1+e^{iax-\varepsilon x}}{1-e^{iax-\varepsilon x}} } 
$&    $(ap,0)$\\ \hline 
         $A_l$ OAM &  $ S_{OAM}(x,y) = \frac{1}{\sqrt{2\pi \varepsilon^2}} e^{-\frac{\left( \tan^{-1}\left(\frac{y}{x}\right)+ax\right)^2}{\varepsilon^2}} $&   $(al,0)$\\ \hline 
 $BG_p$ radial& $ S_{BG}\bigl(x,y\bigr) = \sqrt{\frac{2}{\pi w_0^2}}
  \,e^{-\frac{x^2+y^2}{w_0^2}}
  e^{\frac{2i \left(x^2+y^2\right)}{w_0^2}\,\sin(a x)}$ & $(al,0)$\\ \hline
    \end{tabular}
    \caption{Analytical expressions for the all-mode sorter are provided for several families of spatial modes. Here, $w_0$ denotes the beam waist, and $\varepsilon$ is a small positive constant introduced to prevent numerical instability. {\color{black} The OAM type includes Laguerre–Gaussian and Bessel-Gaussian modes, with orbital angular momentum.}}
    \label{tab:infinity_sorters}
\end{table*}

\section{Discussion}
\label{sec:Discussion}

Our proposed sorter demonstrates notable sorting performance across a broad range of spatial mode families, ideally achieving zero cross-talk. We have shown, both through simulations and experiments, that the sorter performs robustly with various mode types, including Hermite-Gaussian, Laguerre-Gaussian, Bessel-Gaussian, and their combination. For certain mode sets - such as Bessel-Gaussian modes with varying radial indices and mutually unbiased bases (MUBs) - an increase in cross-talk is observed, which aligns with theoretical expectations, as these modes are not orthogonal (see Section~\ref{sub:TheoreticalModel}).

Interestingly, when applied to {\color{black} OAM} modes, the sorter configuration reduces to the well-known Fork sorter, thereby validating the generality and consistency of our approach.

We observed that the sorter exhibits a wavelength sensitivity that increases with the sorting angle: larger sorting angles lead to greater sensitivity to wavelength variations. This effect can be leveraged for spectroscopic applications, as discussed in our theoretical derivation of a Rayleigh-like condition for wavelength resolution. When such sensitivity is undesirable, it can be mitigated through straightforward calibration procedures.

In contrast, the sorter shows significantly lower sensitivity to random, independent phase noise, underscoring its robustness in practical environments.

A power transmission analysis revealed only a moderate deviation from the theoretical scaling prediction of $\mathcal{O}(M^{-1})$, where $M$ denotes the number of modes. While the sorter maintains high fidelity and conceptual simplicity, power loss remains its primary scalability limitation.  Despite this limitation, as demonstrated in the Supplementary Material, our sorter achieves optimal performance {\color{black} as discussed in \ref{sub:TheoreticalModel}.}

We also analyzed the limit of an infinite number of modes ($M \rightarrow \infty$) and demonstrated how the sorter can, in principle, accommodate a large - ideally unbounded - set of modes. In practice, however, realizing such scalability requires extremely high spatial resolution, which can introduce significant power loss. Therefore, for applications involving a modest number of modes, we recommend using the discrete implementation as defined in Equation~\ref{eq:sorter}. For large $M$, it may be more practical to adopt the continuous analytical expression.

Although the sorter design involves both amplitude and phase modulation, we found that a phase-only approximation remains feasible {\color{black} in the cases we have examined}. In particular, when sorting spatial modes with similar amplitude distributions, the phase profile becomes the dominant contributor to performance. Surprisingly, even in cases involving general HG modes with significantly varying amplitudes, the phase-only approximation remains valid. A detailed mathematical analysis of this approximation and its associated error is provided in the Supplementary Material.

{\color{black}
The proposed method offers a distinct alternative to conventional Multi-Plane Light Conversion (MPLC). In terms of signal quality, MPLC performance is typically limited by the convergence of numerical optimization algorithms and the precision of inter-plane alignment, resulting in crosstalk levels typically between -15 dB to -25 dB \cite{fontaine2019laguerre, rocha2025self} In contrast, our method exploits an exact analytical transformation, resulting in near-zero crosstalk. From a complexity perspective, the difference is substantial: we propose a solution utilizing only a single layer, thereby avoiding the bulky cascades of multiple planes (typically 4 to 10) required by MPLC \cite{fontaine2019laguerre, rocha2025self} Regarding efficiency, we acknowledge that our single-plane diffraction induces a power transmission scaling as $O(M^{-1})$, which can be significant for a large number of modes. However, MPLC systems also suffer from non-negligible insertion losses (typically -3.0 to -15.0 dB) \cite{fontaine2019laguerre, rocha2025self} Therefore, we believe that our sorter represents a highly effective solution for applications involving a moderate number of spatial modes, prioritizing signal fidelity and architectural simplicity.

It should be noted that although sorting losses are certainly important, they are not a limiting parameter in many applications. Specifically, in classical spatial division multiplexed systems, losses can be partly compensated by optical amplification. Moreover, in QKD system, the dominant limitation typically comes from channel attenuation, which can easily reach 30–40 dB over metropolitan fiber links, in addition to detection inefficiencies \cite{takesue2007quantum,scarani2009security}. In this context, the sorter loss, e.g. 6 dB for 4 modes, is not the primary bottleneck. Since key generation relies only on successfully detected photons, losses reduce the rate but not the security or correctness of the protocol. This reduction can be compensated by higher-efficiency sources or extended measurement times.

We also compared our method to geometric transformations like Log-Polar \cite{Berkhout} and Angular Lens \cite{AngularLens}. These methods are highly efficient (power loss of -1.5 to -3.5 dB) and compact (1–2 layers). However, they are strictly limited to OAM modes and typically exhibit crosstalk levels around -10 to -20 dB due to geometric approximations. 
}

An additional application of the sorter is its operation in reverse - as a mode generator. This configuration enables the simultaneous generation of a complete set of spatial modes, each with equal power. Specifically, given an incident wavefront with total intensity $I$, each generated mode carries a power of $I/M$. This uniform distribution can be advantageous in applications requiring balanced generation of multiple modes.

\section{Conclusion}
\label{sec:Conclusion}

In summary, we have introduced a universal spatial-mode sorter that achieves near-zero cross-talk across a broad range of mode families, including Hermite-Gaussian, Laguerre-Gaussian, Bessel-Gaussian, and their superpositions. Our approach not only enables high-fidelity mode sorting but also functions as an efficient mode generator, providing a compact and easily designable platform suitable for both classical and quantum photonic systems. While the method remains effective for a large number of spatial modes, the primary limitation in scalability arises from power splitting loss.

Future research directions include extending the applicability of our technique to other domains, such as electron beam shaping \cite{verbeeck2012new, mcmorran2012platelet} and spatial-mode manipulation in fiber optics \cite{bozinovic2013terabit}. An additional avenue of interest is enhancing the mode generation aspect to achieve greater efficiency and flexibility. A particularly promising direction is the use of this technique for sorting quantum wavefunctions, for example in quantum key distribution (QKD), where the verification of mutually unbiased basis (MUB) measurements in the classical domain presented here may serve as a foundation.

In this work, the sorter was realized by a spatial light modulator, but in the future, a compact and fixed device can be realized by direct laser printing \cite{lightman2022integrated}.

A code to create sorters is provided in Code 1 (Ref.~\cite{MMScode}).


\begin{backmatter}
\bmsection{Funding} 
K.C. gratefully acknowledges the Milner Foundation for their support. This work was supported by the Israeli Ministry of Innovation, Science and Technology, Grant No. 1001572598; the Tel Aviv University Center for Artificial Intelligence; the Israeli Science Foundation (ISF) Excellence Center Grant No. 2312/21 and ISF Grant No. 969/22; the US-Israel Binational Science Foundation (BSF).

\bmsection{Acknowledgment} 
The authors thank Joseph Meyer for his assistance in building the experimental setup during the early stages of the project.

\bmsection{Disclosures} The authors declare no conflicts of interest.

\bmsection{Data availability} Data underlying the results presented in this paper can be easily generated using the shared project code.

\bmsection{Supplemental document}
See Supplementary document for supporting the material in the paper.
\end{backmatter}

\bibliography{references}

\newpage
\appendix

\section{Sorter Math derivation}
\label{app:math_derivation}
{\color{black}In this section, we describe the mathematical formulation of the setup and define the sorter}. We further show mathematically why our suggested sorter works well and {\color{black}discuss its limitations}.

We start with the expression of the far field (Fraunhofer approximation):
\begin{gather}
    E_m(\xi, \eta) = FT\left\{ f_m(x,y) S(x,y) \right\}_{\nu_x =\frac{\xi}{\lambda z}, \nu_y=\frac{\eta}{ \lambda z}}  \ ,
\end{gather}
\begin{gather}
    E_m(\xi, \eta) = \int \int f_m(x,y) S(x,y) e^{-i\frac{2\pi}{\lambda z}\left( \xi x + \eta y\right) } dx dy \ .
\end{gather}
We are interested in detecting the wavefront centered around some detector $\mu$ located at $ d_{\mu} = (\alpha_\mu, \beta_\mu)$. Since we are interested in a point detector, we substitute the value at $\xi = \alpha_\mu$, and $\eta = \beta_\mu$:
\begin{gather}
    E_{m,\mu}= \int \int f_m(x,y) S(x,y) e^{-i\frac{2\pi}{\lambda z}\left( \alpha_\mu x + \beta_\mu y\right) } dx dy \ ,
\end{gather}
Or:
\begin{gather}
    E_{m, \mu} = FT\left\{ f_m(x,y) S(x,y) \right\}_{\nu_x =\frac{\alpha_\mu}{\lambda z}, \nu_y=\frac{\beta_\mu}{ \lambda z}} \ .
\end{gather}
We want to maximize this value as much as possible for $m = \mu$ and minimize for $m\neq \mu$.

We write $S(x,y)$ as a sum of some functions, in the following form:
\begin{gather}
    S(x,y) \equiv \frac{1}{\sqrt{M}}\sum_{m^\prime} C^{*}_{m^\prime}(x,y) e^{i\frac{2\pi}{\lambda z}\left( \alpha_{m^\prime} x + \beta_{m^\prime} y\right) } \ .
\end{gather}
{\color{black}Substituting the sorter into the expression yields}:
\begin{gather}
    E_{m,\mu}= \int \int f_m(x,y) \left[\frac{1}{\sqrt{M}} \sum_{m^\prime} C^{*}_{m^\prime}(x,y) e^{i\frac{2\pi}{\lambda z}\left( \alpha_{m^\prime} x + \beta_{m^\prime} y\right) } \right] e^{-i\frac{2\pi}{\lambda z}\left( \alpha_\mu x + \beta_\mu y\right) } dx dy \\ \nonumber
= \frac{1}{\sqrt{M}}\sum_{m^\prime} \int \int f_m(x,y)  C^{*}_{m^\prime}(x,y) e^{i\frac{2\pi}{\lambda z}\left[ (\alpha_{m^\prime}-\alpha_\mu) x + (\beta_{m^\prime}-\beta_\mu) y\right] }  dx dy \ .
\end{gather}
Separate the sum into the diagonal and the off-diagonal:
\begin{gather}
    E_{m,\mu}= \frac{1}{\sqrt{M}}\int \int  C^{*}_{\mu}(x,y) f_m(x,y)  dx dy \\ \nonumber
  +  \frac{1}{\sqrt{M}}\sum_{m^\prime \neq \mu} \int \int C^{*}_{m^\prime}(x,y) f_m(x,y)  e^{i\frac{2\pi}{\lambda z}\left[ (\alpha_{m^\prime}-\alpha_\mu) x + (\beta_{m^\prime}-\beta_\mu) y\right] }  dx dy \ .
\end{gather}
The first term is the inner product between $C_\mu (x,y)$ and $f_m(x,y)$ and the second term can be considered as a Fourier transform of the multiplication of the two functions, evaluated at a particular frequency:
\begin{gather}
    E_{m,\mu} = \frac{1}{\sqrt{M}}\langle {C_\mu} | {f_m}\rangle + \frac{1}{\sqrt{M}} \sum_{m^{\prime} \neq  \mu} FT\left\{ C^*_{m^\prime} f_m  \right\}_{\nu_x=\alpha_{m^\prime}-\alpha_\mu, \nu_y = \beta_{m^\prime}-\beta_\mu} \ .
\end{gather}
Choosing $C_\mu(x,y) = f_m(x,y)$ leads to:
\begin{gather}
\label{eq:sorter}
    S(x,y) = \frac{1}{\sqrt{M}} \sum_{m^\prime=1}^M f_{m^\prime}^*(x,y) e^{i\frac{2\pi}{\lambda z} \left( \alpha_{m^{\prime}}x +\beta_{m^\prime}y \right)} \ ,
\end{gather}
and:
\begin{gather}
    E_{m,\mu} = \frac{1}{\sqrt{M}}\langle {f_\mu} | {f_m}\rangle + \frac{1}{\sqrt{M}} \sum_{m^{\prime} \neq  \mu} FT\left\{ f^*_{m^\prime} f_m  \right\}_{\nu_x=\alpha_{m^\prime}-\alpha_\mu, \nu_y = \beta_{m^\prime}-\beta_\mu} \ .
\end{gather}
We consider the first term as the signal, the the second term as the noise. For a general set of modes, typically, it turns out that the second term is very small, and we get approximately:
\begin{gather}
    E_{m,\mu} \approx \frac{1}{\sqrt{M}}\langle {f_\mu} | {f_m}\rangle \\ 
    I_{m,\mu} \approx \frac{1}{M}\left| \langle {f_\mu} | {f_m}\rangle \right|^2  \ .
\end{gather}
Evaluating the cross-talk over all the modes:
\begin{gather}
    CT = \frac{\sum_{\mu > m} \left| E_{m,\mu} \right|^2 }{\sum_{m} \left| E_{m,m} \right|^2 }= \frac{1}{M} \sum_{\mu > m} \left|\langle {f_\mu} | {f_m}\rangle \right|^2 \ .
\end{gather}

And if we take orthonormal set of modes:
\begin{gather}
    E_{m,\mu} = \frac{1}{\sqrt{M}}\delta_{m,\mu} + \frac{1}{\sqrt{M}} \sum_{m^{\prime} \neq  \mu} FT\left\{ f^*_{m^\prime} f_m  \right\}_{\nu_x=\alpha_{m^\prime}-\alpha_\mu, \nu_y = \beta_{m^\prime}-\beta_\mu} \ .
\end{gather}

The size of the scattering area around each of the detectors is given by the bandwidth of the term $f_{m^\prime}^*f_m$ and bounded by the maximal bandwidth among all the modes. We define this maximum bandwidth as $\Omega$ and it defines the minimum separation between two detectors to avoid any crosstalk:
\begin{gather}
    \forall \mu, m \ , \ \sqrt{ (\alpha_\mu - \alpha_m)^2+(\beta_\mu - \beta_m)^2   }\geq \Omega \ .
\end{gather}

Since the Fourier transform of a Gaussian function with standard deviation $\sigma$ is another Gaussian with standard deviation $\frac{1}{\sigma}$, we can approximate the bandwidth of the beam accordingly. For a Gaussian beam with a given $M^2$ factor, the effective beam width is approximately $w M$. Thus, the corresponding angular bandwidth can be estimated as $\Omega \approx \frac{\lambda z}{2w \bar{M}}$, where $w$ is the width of the incident beam, $z$ is the propagation distance (far field), $\lambda$ is the wavelength, and $\bar{M}$ is the minimal $M$ factor (the square-root of the $M^2$ factor) across all modes.
    
So we conclude that there are four key findings:
\begin{enumerate}
    \item The intensity transmission coefficient goes as $\sim O(M^{-1})$
    \item The crosstalk depends on the square average of the inner-products between the modes
    \item Typically, for orthogonal modes, the cross-talk is zero
    \item The separation should be larger than the bandwidth which can be approximated using M-squared.
\end{enumerate}

\newpage

\section{Sorter Optimality for one layer - a proof}
\label{app:sorter_optimality}

Given some arbitrary sorter $S(x)$, we express the output field as:
\begin{gather}
    E_{m, \mu} = FT\left\{ f_m(x,y) S(x,y) \right\}_{\nu_x =\frac{\alpha_\mu}{\lambda z}, \nu_y=\frac{\beta_\mu}{ \lambda z}} \ .
\end{gather}
We want to maximize this value as much as possible for $m = \mu$ and minimize for $m\neq \mu$. Focusing on the case where $m = \mu$:
\begin{gather}
    I_{m}(\alpha_m, \beta_m) \equiv \left| FT\left\{ f_m(x,y) S(x,y) \right\}_{\nu_x =\frac{\alpha_m}{\lambda z}, \nu_y=\frac{\beta_m}{ \lambda z}} \right|^2 \\ \nonumber
= \int_{-\infty}^{\infty}\int_{-\infty}^{\infty} f_m^*(x',y') S^*(x',y') e^{i\frac{2\pi}{\lambda z}\left( \alpha_m x' + \beta_m y'\right) } dx' dy'
    \int_{-\infty}^{\infty}\int_{-\infty}^{\infty} f_m(x,y) S(x,y) e^{-i\frac{2\pi}{\lambda z}\left( \alpha_m x + \beta_m y\right) } dx dy \\ \nonumber
   = \int_{-\infty}^{\infty}\int_{-\infty}^{\infty}\int_{-\infty}^{\infty}\int_{-\infty}^{\infty} f_m^*(x',y') f_m(x,y)  S^*(x',y')  S(x,y) e^{i\frac{2\pi}{\lambda z}\Big[ \alpha_m (x'-x) + \beta_m (y'-y)\Big] } 
     dx' dy' dx dy \ .
\end{gather}

The average transmission power is given by: 
\begin{gather} \label{eq:transmission_coeff}
    \mathcal{T} = \frac{1}{M}\sum_{m=1}^{M} I_m (\alpha_m, \beta_m) \ .
\end{gather}
We set the detectors location to be random variables, to model the fact that the detectors are located in an arbitrary location, and we are interested in the expectation value of the transmission power $\mathcal{T}$:
\begin{gather}
    \mathbf{E}_{ \{\alpha,\beta\} }[ \mathcal{T} ] = \frac{1}{M}\sum_{m=1}^{M} \mathbf{E}_{\{\alpha,\beta\}}\Big[ I_m (\alpha_m, \beta_m) \Big] \ .
\end{gather}
The expectation value over a single mode $m$:
\begin{gather}
        \mathbf{E}_{\{\alpha,\beta\}}\Big[ I_m (\alpha_m, \beta_m) \Big] = \int_{-\infty}^{\infty}\int_{-\infty}^{\infty}\int_{-\infty}^{\infty}\int_{-\infty}^{\infty} f_m^*(x',y') f_m(x,y) \\ \nonumber \cdot  \   S^*(x',y')  S(x,y)\mathbf{E}_{\{\alpha,\beta\}}\Big[ e^{i\frac{2\pi}{\lambda z}\Big[ \alpha_m (x'-x) + \beta_m (y'-y)\Big] } \Big]
     dx' dy' dx dy \ .
\end{gather}
We assume here that the detectors {\color{black}have} an equal probability to be located in any arbitrary location, therefore, we model their probability density function to be a Gaussian with some large enough standard deviation $\sigma$. One could notice that the expectation value is exactly the characteristic function of the distribution, evaluated at a particular location $\Big(\frac{2\pi}{\lambda z}(x'-x), \frac{2\pi}{\lambda z}(y'-y)\Big)$. Therefore:
\begin{gather}
    p(\alpha_m, \beta_m) =  p(\alpha_m) p(\beta_m) = \frac{1}{2\pi\sigma^2} e^{-\frac{\alpha_m^2 + \beta_m^2}{2\sigma^2}},
\end{gather}
And using the characteristic function of a Gaussian distribution:
\begin{gather}
    \mathbf{E}_{\{\alpha,\beta\}}\Big[ e^{i\frac{2\pi}{\lambda z}\Big[ \alpha_m (x'-x) + \beta_m (y'-y)\Big] } \Big] = e^{-\frac{2\pi^2\sigma^2}{\lambda^2 z^2} \Big[ (x - x')^2 + (y-y')^2 \Big] } \ .
\end{gather}

For a sufficiently large distance of detectors from the optical axis, i.e. $\sigma \gg \lambda z$, which is a very typical scenario, we can approximate this expectation value to be a delta function:
\begin{gather}
    \lim_{\sigma \rightarrow \infty} \mathbf{E}_{\{\alpha,\beta\}}\Big[ e^{i\frac{2\pi}{\lambda z}\Big[ \alpha_m (x'-x) + \beta_m (y'-y)\Big] } \Big] = \delta(x-x') \delta(y-y') \ .
\end{gather}
The final result for the average power transmission is given by:
\begin{gather}
    \mathbf{E}_{ \{\alpha,\beta\} }[ \mathcal{T} ] = \frac{1}{M}\sum_{m=1}^{M}  \int_{-\infty}^{\infty}\int_{-\infty}^{\infty} |f_m(x,y)S(x,y)|^2  dx dy \ .
\end{gather}
Rearranging the terms:
\begin{gather}
    \mathbf{E}_{ \{\alpha,\beta\} }[ \mathcal{T} ] = \int_{-\infty}^{\infty}\int_{-\infty}^{\infty} \underbrace{\bigg[ \frac{1}{M}\sum_{m=1}^{M} f_m^*(x,y) f_m(x,y) \bigg]}_{\Phi_1}  \underbrace{\bigg[ S^*(x,y)  S(x,y) \bigg] }_{\Phi_2}  dx dy \ .
\end{gather}
For bounded energy of the functions $f_m(x,y)$ and the sorter $S(x,y)$, this integral will be maximized when the $\Phi_1$ is fully aligned with $\Phi_2$, in other words, the optimal sorter $\hat{S}(x,y)$ satisfied the following condition:
\begin{gather}
    \hat{S}^*(x,y) \hat{S}(x,y) = \frac{1}{M}\sum_{m=1}^{M} f_m^*(x,y) f_m(x,y) \ ,
\end{gather}
for any coordinate $x,y$.
This relation is correct also for any phase multiplication:
\begin{gather}
\label{eq:optimalSorter1}
    \hat{S}^*(x,y) \hat{S}(x,y) = \frac{1}{M}\sum_{m=1}^{M} f_m^*(x,y)e^{i\frac{2\pi}{\lambda z} \left( \alpha_{m}x +\beta_{m}y \right)} f_m(x,y)e^{-i\frac{2\pi}{\lambda z} \left( \alpha_{m}x +\beta_{m}y \right)} \ ,
\end{gather}
for any specific choice of detector location $\alpha_m, \beta_m$.
Since, we argue that the functions are well separated in the frequency domain (since they are band-pass limited), we can follow the same argument from the previous section to state that:
\begin{gather}
\label{eq:optimalSorter2}
   \int_{-\infty}^{\infty}\int_{-\infty}^{\infty} f_m^*(x,y) f_m(x,y) f_n^*(x,y)f_{n'}(x,y)e^{-i\frac{2\pi}{\lambda z} \left[ (\alpha_{n} - \alpha_{n'})x +(\beta_{n} - \beta_{n'})y \right]} dx dy \\ \nonumber
   = FT\left\{ \left| f_m(x,y) \right|^2 f_n^*(x,y)f_{n'} \right\}_{\nu_x=\alpha_{n^\prime}-\alpha_{n}, \nu_y = \beta_{n^\prime}-\beta_{n}} \approx 0 \ ,
\end{gather}
for any $n \neq n^\prime$.
From this reason, we can write the condition for the optimal sorter (eq. \ref{eq:optimalSorter1}) as follows:
\begin{gather}
    \hat{S}^*(x,y) \hat{S}(x,y) = \Bigg[ \frac{1}{\sqrt{M}}\sum_{m=1}^{M} f_m(x,y)e^{-i\frac{2\pi}{\lambda z} \left( \alpha_{m}x +\beta_{m}y \right)} \Bigg]^* \Bigg[  \frac{1}{\sqrt{M}}\sum_{m=1}^{M} f_m(x,y)^* e^{i\frac{2\pi}{\lambda z} \left( \alpha_{m}x +\beta_{m}y \right)} \Bigg] \ .
\end{gather}
Because the excess terms do not contribute, following eq. \ref{eq:optimalSorter2}.
Finally, we get:
\begin{gather}
    \hat{S}(x,y) = \frac{1}{\sqrt{M}}\sum_{m=1}^{M} f_m(x,y)^* e^{i\frac{2\pi}{\lambda z} \left( \alpha_{m}x +\beta_{m}y \right)} \ ,
\end{gather}
which is exactly our suggested sorter.

\begin{figure}
    \centering
    \includegraphics[width=1.0\linewidth]{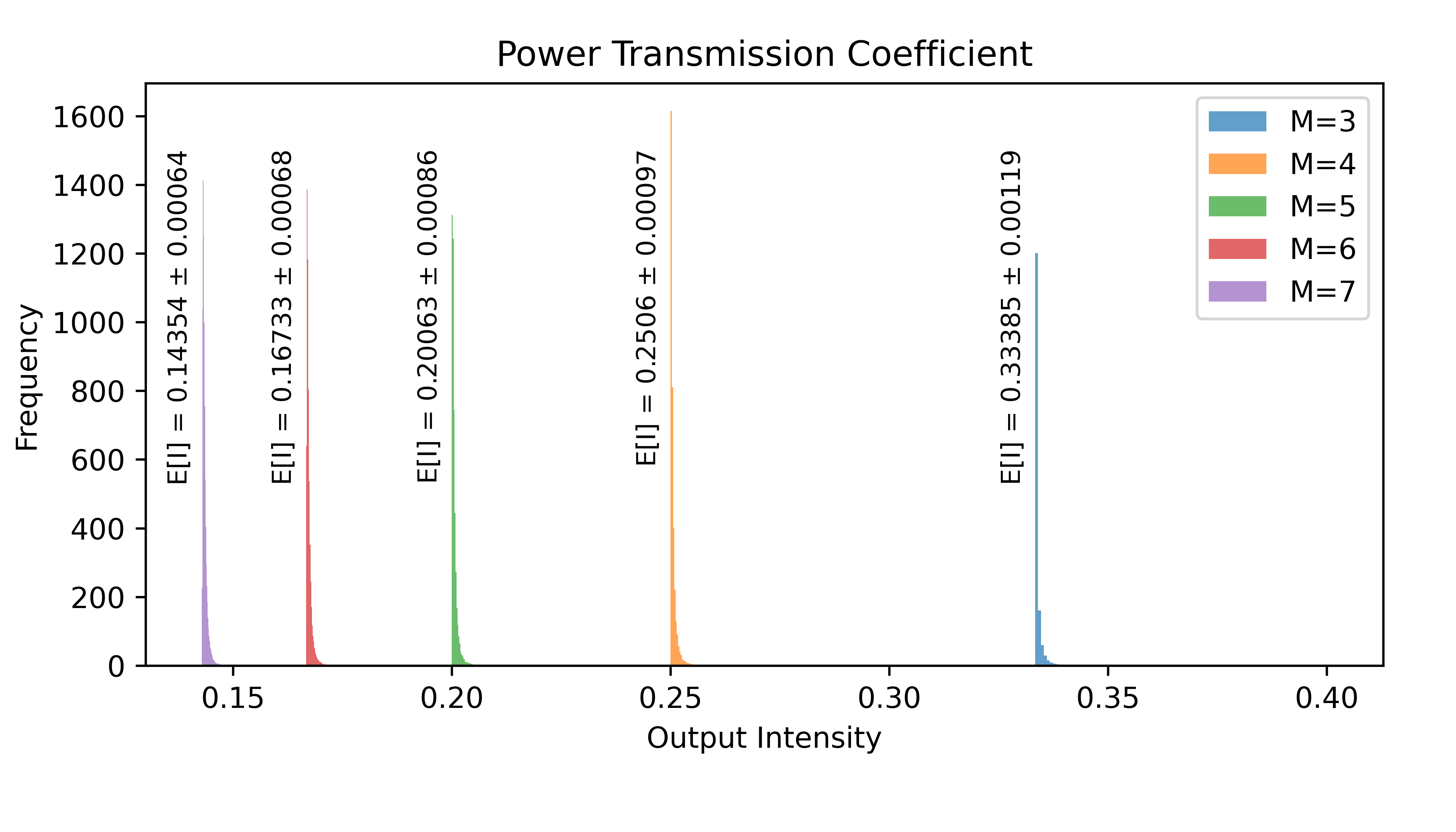}
    \caption{{\color{black} Random sampling of different detector positions for HG and LG random modes (up to mode 10). Each histogram contains 50,000 randomly chosen detector positions and mode families to sort. The detectors were placed randomly following a normal distribution around the optical axis, ensuring good coverage of the entire output plane.}}
    \label{fig:powerConcentration}
\end{figure}

\begin{figure}
    \centering
    \includegraphics[width=1.0\linewidth]{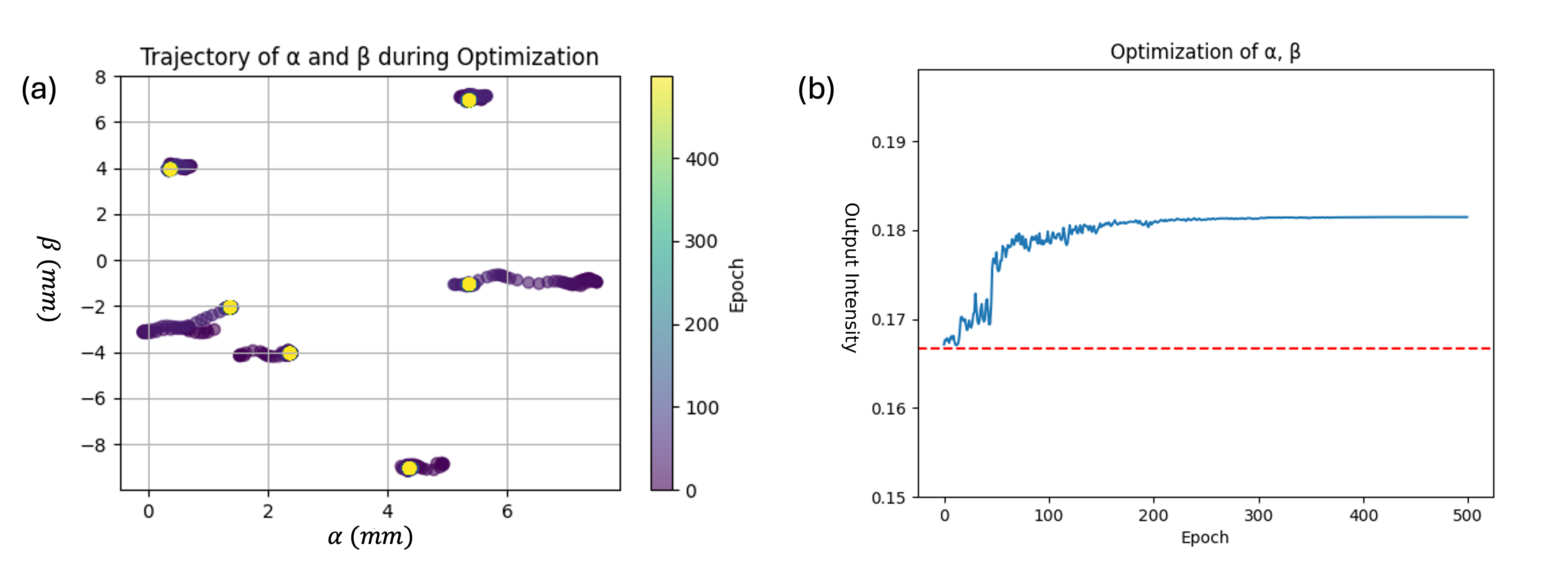}
    \caption{{\color{black} Optimization over the detectors location, can still improve the results but this is very limited. An example of an optimization over HG modes: $HG_{00},HG_{01},HG_{10},HG_{11},HG_{02},HG_{20}$. Left: Evolution of the detector locations over the optimization epochs. Right: Output intensity coefficient as a function of the optimization epoch; the red horizontal dashed line indicates the $M^{-1}$ baseline.}}
    \label{fig:powerOptimization}
\end{figure}

{\color{black} Empirical observations show that the variance of the power transmission coefficient (eq. \ref{eq:transmission_coeff}) concentrates around its mean value, as shown in Figure \ref{fig:powerConcentration}. Concretely, the variance of the power transmission are in the order of $10^{-7}$.

Moreover, we additionally demonstrate that optimizing the detector position to maximize the overall transmission coefficient still offers only very limited improvement, this is shown in Figure \ref{fig:powerOptimization}.

Additional evidence from the numeric indicate that the total power transmission is bounded from below by $M^{-1}$.
}

\newpage

\section{Extension to an infinite number of modes}
\label{app:ext2infmodes}

{\color{black}In this section, we derive closed-form analytical expressions} for all HG modes, LG modes and BG modes using their generating functions.

\subsection{Hermite-Gaussian}
The normalized two-dimensional Hermite-Gaussian modes are given by the following expression:
\begin{equation}
\mathrm{HG}_{n,m}(x,y) = \sqrt{\frac{2}{\pi}} \, \frac{1}{w_0} \, \frac{1}{\sqrt{2^{n+m} n! m!}} \, H_n\left(\frac{\sqrt{2}x}{w_0}\right) H_m\left(\frac{\sqrt{2}y}{w_0}\right) e^{-\frac{x^2 + y^2}{w_0^2}} \ ,
\end{equation}
where $H_n(\cdot)$ denotes the physicists' Hermite polynomials, and $w_0$ is the beam waist.

The generating function for the Hermite polynomials in one variable is
\begin{equation}
\sum_{n=0}^{\infty} \frac{t^n}{n!} H_n(x) = \exp\left(2xt - t^2\right) \ ,
\end{equation}
where $t$ is an auxiliary parameter.

Therefore, the two-dimensional generating function for the HG modes becomes
\begin{align}
S_{\mathrm{HG}}^*(x,y; t_x, t_y)&= \sum_{n=0}^{\infty} \sum_{m=0}^{\infty} \frac{t_x^n t_y^m}{\sqrt{2^{n+m} n! m!}} \, \mathrm{HG}_{n,m}(x,y) \\ \nonumber
&= \sqrt{\frac{2}{\pi}} \, \frac{1}{w_0} \, e^{-\frac{x^2 + y^2}{w_0^2}} \sum_{n=0}^{\infty} \frac{(t_x/\sqrt{2})^n}{n!} H_n\left(\frac{\sqrt{2}x}{w_0}\right) \sum_{m=0}^{\infty} \frac{(t_y/\sqrt{2})^m}{m!} H_m\left(\frac{\sqrt{2}y}{w_0}\right) \ .
\end{align}
Using the one-dimensional generating function twice (once for $x$ and once for $y$), we find
\begin{equation}
S_{\mathrm{HG}}^*(x,y; t_x, t_y) = \sqrt{\frac{2}{\pi}} \, \frac{1}{w_0} \, \exp\left( -\frac{x^2 + y^2}{w_0^2} + \frac{2}{w_0} (x t_x + y t_y) - (t_x^2 + t_y^2) \right) \ .
\end{equation}
Thus, the generating function for the 2D Hermite--Gaussian modes is a simple Gaussian in $(x,y,t_x,t_y)$.

Taking: $t_x = e^{-iax}, t_y = e^{-iby}$ to get the final:
\begin{equation}
S_{\mathrm{HG}}(x,y) = \sqrt{\frac{2}{\pi}} \, \frac{1}{w_0} \, \exp\left( -\frac{x^2 + y^2}{w_0^2} + \frac{2}{w_0} (x e^{-iax} + y e^{-iby}) - e^{-2iax} - e^{-2iby} \right) \ .
\end{equation}

\subsection{Laguerre-Gaussian with radial index only}

For the Laguerre-Gaussian (LG) modes with azimuthal index $l=0$, the radial envelope is given by
\begin{equation}
\mathrm{LG}_p(r) = \sqrt{\frac{2}{\pi}} \, \frac{1}{w_0} \, L_p\left( \frac{2r^2}{w_0^2} \right) e^{-\frac{r^2}{w_0^2}} \ ,
\end{equation}
where $r$ is the radial index , and $L_p(\cdot)$ are the Laguerre polynomials.

The generating function for the Laguerre polynomials is given by:
\begin{equation}
\sum_{p=0}^{\infty} t^p L_p(x) = \frac{1}{1-t} \exp\left( -\frac{t}{1-t} x \right) \ ,
\quad \text{for} \quad |t| < 1 \ .
\end{equation}

Thus, the generating function for the Laguerre-Gaussian modes becomes
\begin{align}
S^*_{\mathrm{LG}}(r; t) &= \sum_{p=0}^{\infty} t^p \, \mathrm{LG}_p(r) \\ \nonumber
&= \sqrt{\frac{2}{\pi}} \, \frac{1}{w_0} \, e^{-\frac{r^2}{w_0^2}} \sum_{p=0}^{\infty} t^p L_p\left( \frac{2r^2}{w_0^2} \right) \\ \nonumber
&= \sqrt{\frac{2}{\pi}} \, \frac{1}{w_0} \, e^{-\frac{r^2}{w_0^2}} \, \frac{1}{1-t} \, \exp\left( -\frac{2s r^2 / w_0^2}{1-t} \right) \ .
\end{align}
Simplifying the exponents, we obtain
\begin{equation}
S^*_{\mathrm{LG}}(r; t) = \sqrt{\frac{2}{\pi}} \, \frac{1}{w_0 (1-t)} \exp\left( -\frac{r^2}{w_0^2} \frac{1+t}{1-t} \right) \ .
\end{equation}
One may notice that the requirement $|t|<1$ means we can not simply substitute $t=e^{-iax}$, but we need to artificially add some small attenuation: $t = e^{-iax -\varepsilon x}$:
\begin{equation}
S_{\mathrm{LG}}(r,x) = \sqrt{\frac{2}{\pi}} \, \frac{1}{w_0 (1-e^{iax-\varepsilon x})} \exp\left( -\frac{r^2}{w_0^2} \frac{1+e^{iax-\varepsilon x}}{1-e^{iax-\varepsilon x}} \right) \ .
\end{equation}

\subsection{OAM}

Say we want to sort only the angular momentum part of the mode::
\begin{gather}
    A_l(\phi) = e^{i\phi l} \ .
\end{gather}
The generating function can be formulated as:
\begin{gather}
    \sum_{l=-\infty}^\infty A_l(\phi) t^l \ .
\end{gather}
Plugging $t=e^{ixa}$:
\begin{gather}
    S^*_{OAM}(\phi,x) = \sum_{l=-\infty}^\infty e^{i\phi l } \left( e^{ixa} \right)^l = \delta(\phi + a x) \approx \frac{1}{\sqrt{2\pi \varepsilon^2}} e^{-\frac{(\phi+ax)^2}{\varepsilon^2}} \ ,
\end{gather}
While in the last step we used the fact that a delta function can be approximated using a Gaussian distribution with standard deviation goes to 0.
Therefore, the sorter is given by:
\begin{equation}
S_{OAM}(\phi,x) = \frac{1}{\sqrt{2\pi \varepsilon^2}} e^{-\frac{(\phi+ax)^2}{\varepsilon^2}}.
\end{equation}

\subsection{Bessel–Gaussian without OAM}

We define the family of normalized Bessel–Gaussian modes (order $p$):
\begin{equation}
\mathrm{BG}_p(r)
= \sqrt{\frac{2}{\pi\,w_0^2}}
\;J_p\!\Bigl(\tfrac{2r^2}{w_0^2}\Bigr)
\;\exp\!\bigl(-\tfrac{r^2}{w_0^2}\bigr) \ ,
\end{equation}
where $J_p$ is the Bessel function of order $p$.

The generating‐function identity for $J_p$ is
\begin{equation}
\sum_{p=-\infty}^{\infty} t^p\,J_p(r)
= \exp^{\frac{r}{2}\bigl(t - t^{-1}\bigr)} \ .
\end{equation}
{\color{black}From} the BG generating function,
\begin{align}
S^*_{BG}(r;t)
&= \sum_{p=-\infty}^\infty t^p\,\mathrm{BG}_p(r) \\ \nonumber
&= \sqrt{\frac{2}{\pi w_0^2}}
  \,e^{-r^2/w_0^2}
  \sum_{p=-\infty}^\infty t^p
      J_p\!\Bigl(\tfrac{2r^2}{w_0^2}\Bigr) \\ \nonumber
&= \sqrt{\frac{2}{\pi w_0^2}}
  \,e^{-r^2/w_0^2}
  \exp\!\Bigl[\tfrac{1}{2}\,\tfrac{2r^2}{w_0^2}\,(t - t^{-1})\Bigr] \\ \nonumber
&= \sqrt{\frac{2}{\pi w_0^2}}
  \,e^{-r^2/w_0^2}
  \exp\!\Bigl[\tfrac{r^2}{w_0^2}\,(t - t^{-1})\Bigr] \ .
\end{align}
{\color{black}Plugging} the periodic exponent and the small attenuation as before: $t = e^{-i a x}$ .
Then
\begin{gather}
t - t^{-1}
= e^{-i a x} - e^{i a x}
= -2i\,\sin(a x) \ .
\end{gather}
Hence the propagated field is
\begin{align}
S^*_{BG}\bigl(r\bigr)
&= \sqrt{\frac{2}{\pi w_0^2}}
  \,e^{-r^2/w_0^2}
  \exp\!\Bigl[\tfrac{r^2}{w_0^2}\,(t - t^{-1})\Bigr] \\ \nonumber
&= \sqrt{\frac{2}{\pi w_0^2}}
  \,e^{-r^2/w_0^2}
  \exp\!\Bigl[-2\,i\,\tfrac{r^2}{w_0^2}\,\sin(a x)\Bigr] \ .
\end{align}
If one further rescales units so that $2r^2/w_0^2\to1$, this reduces to
\begin{equation}
S_{BG}\bigl(r,x\bigr) = \sqrt{\frac{2}{\pi w_0^2}}
  \,e^{-r^2/w_0^2}
  e^{2i\frac{r^2}{w_0^2}\,\sin(a x)} \ .
\end{equation}

\newpage

\section{Projection to Phase-only Sorter}
\label{app:phase_only}

Given a complex field $S(x,y) = A_s(x,y) e^{i\theta_s(x,y)}$, we seek the phase-only sorter $e^{i\phi(x,y)}$ that minimizes the difference between the projected field for each of the modes (output wave-field):
\begin{gather}
    \forall m, \mu \ \min { \left| \int dx dy \left( S(x,y)-e^{i\phi(x,y) }\right) f_m(x,y) e^{i(\nu_x x + \nu_y y)} \right|^2 } \ .
\end{gather}
This can be bounded by the requirement:
\begin{equation}
    \min_{\phi(x,y)} \left| S(x,y) - e^{i\phi(x,y)} \right|^2 \ .
    \label{eq:min_phase}
\end{equation}

We define $A_s(x,y) = |S(x,y)|$ and $\theta_s(x,y) = \arg S(x,y)$, and expanding the squared magnitude to get:
\begin{align}
    \left| A_s e^{i\theta_s} - e^{i\phi} \right|^2 
    = \left| A_s\cos\theta_s - \cos\phi \right|^2 + \left| A_s\sin\theta_s - \sin\phi \right|^2  
    = A_s^2 + 1 - 2A_s\cos(\theta_s - \phi) \ ,  
    \label{eq:cost_expanded}
\end{align}
using the identity $\cos(\theta_s - \phi) = \cos\theta_s\cos\phi + \sin\theta_s\sin\phi$.

Minimizing by taking the functional derivative with respect to $\phi(x,y)$:
\begin{align}
    \frac{\delta}{\delta\phi} \left[ A_s^2 + 1 - 2A_s\cos(\theta_s - \phi) \right] = 2A_s\sin(\theta_s - \phi) \ .
    \label{eq:functional_derivative}
\end{align}
equating to zero, to get:
\begin{equation}
    \sin(\theta_s - \phi) = 0 \implies \phi = \theta_s + k\pi,\quad k \in \mathbb{Z} \ .
    \label{eq:critical_points}
\end{equation}
Substitute $\phi = \theta_s + k\pi$ back into the cost function \ref{eq:min_phase}:
\begin{equation}
    \left| S - e^{i\phi} \right|^2 = 
    \begin{cases}
        (A_s - 1)^2, & k \text{ even} \\ \nonumber
        (A_s + 1)^2, & k \text{ odd} \ .
    \end{cases}
\end{equation}
The minimum occurs when $k$ is even (i.e., $\phi = \theta_s$), provided $A_s \geq 0$ (always true by definition).
Namely, taking the $\phi(x,y) = \arg S(x,y)$ is the best approximation for the output field given by the original sorter $S(x,y)$, over any given input.

{\color{black}The error can then be evaluated as}: $e = |S(x,y)-1|^2$, which vanishes when $|S(x,y)| = 1$ (when the sorter is already a phase-only mask).

\begin{figure}
    \centering
    \includegraphics[width=1.0\linewidth]{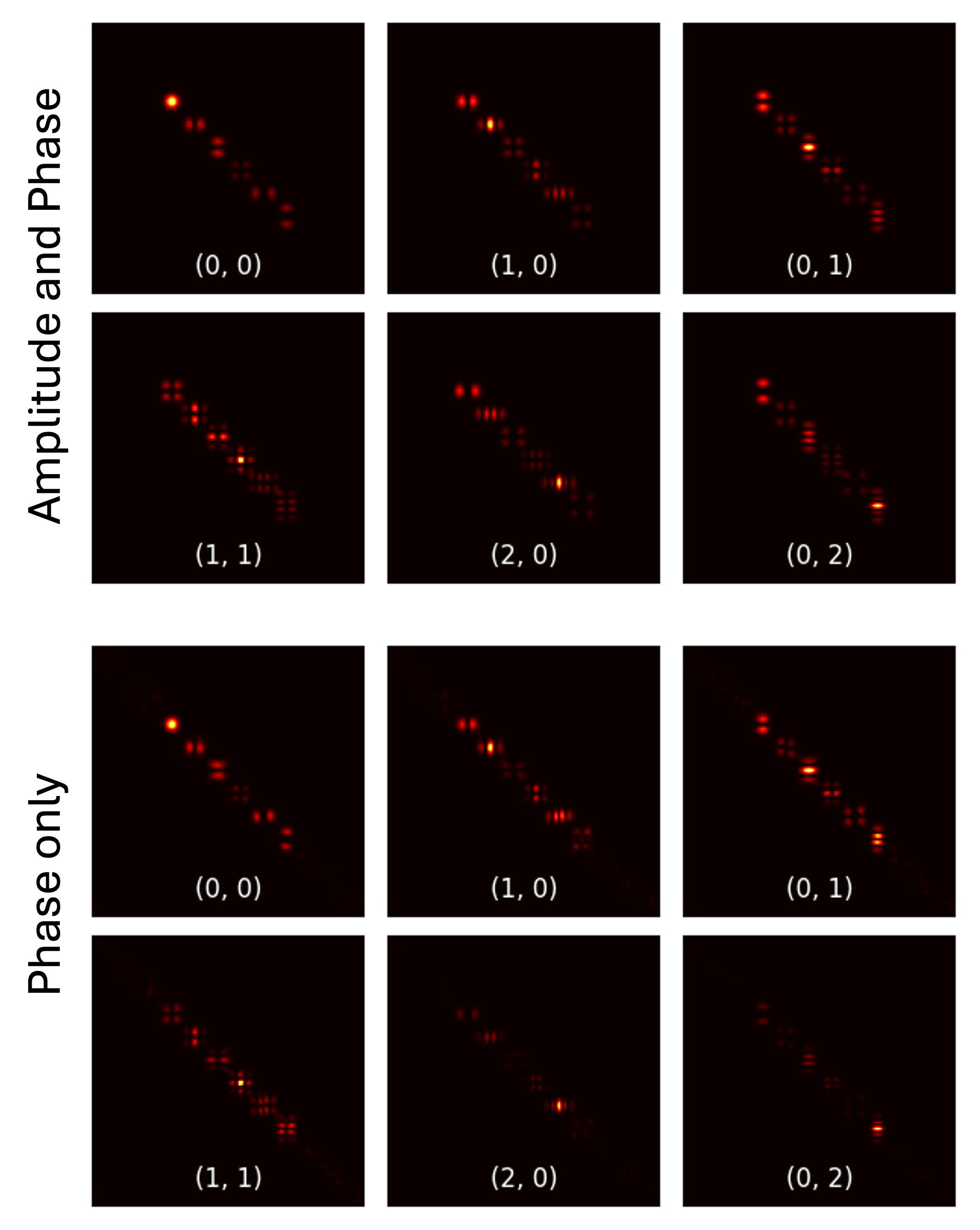}
    \caption{Simulation results showing the projection of the true sorter (top), along with the phase-only approximation (bottom). The sorted modes are $HG_{0,0},HG_{1,0},HG_{0,1}, HG_{1,1},HG_{0,2}, HG_{2,0}$. }
    \label{fig:phase_only_analysis}
\end{figure}

\begin{figure}
    \centering
    \includegraphics[width=1.0\linewidth]{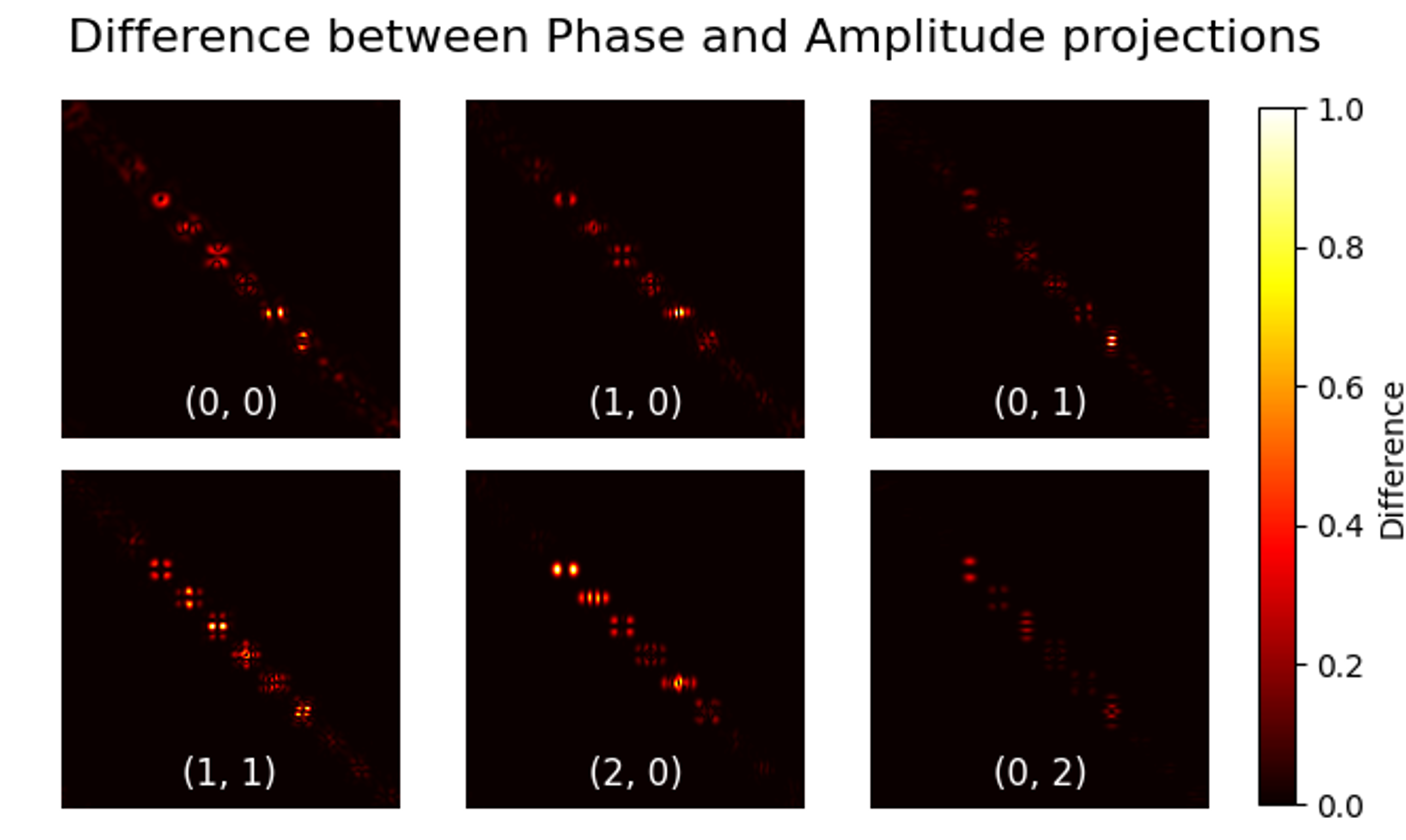}
    \caption{Simulation results of the intensities differences between the projection of the true sorter, and the phase-only approximation. The peak intensities were normalized to unit., and the sorted modes are $HG_{0,0},HG_{1,0},HG_{0,1}, HG_{1,1},HG_{0,2}, HG_{2,0}$. }
    \label{fig:diff_phase_only_analysis}
\end{figure}

Figure \ref{fig:phase_only_analysis} shows a comparison between the projections of Hermite-Gaussian modes for the true sorter, and its phase-only approximator. Figure \ref{fig:diff_phase_only_analysis} shows the absolute value of their difference, after normalizing the peak intensity to unit. {\color{black} Evidently, we observed no substantial differences in the projection patterns and the confusion matrices between the two cases, in both the simulations and the experiments.}

\newpage

\section{Mix different Modes}
\label{app:mix_modes}
\begin{figure}
    \centering
    \includegraphics[width=1.0\linewidth]{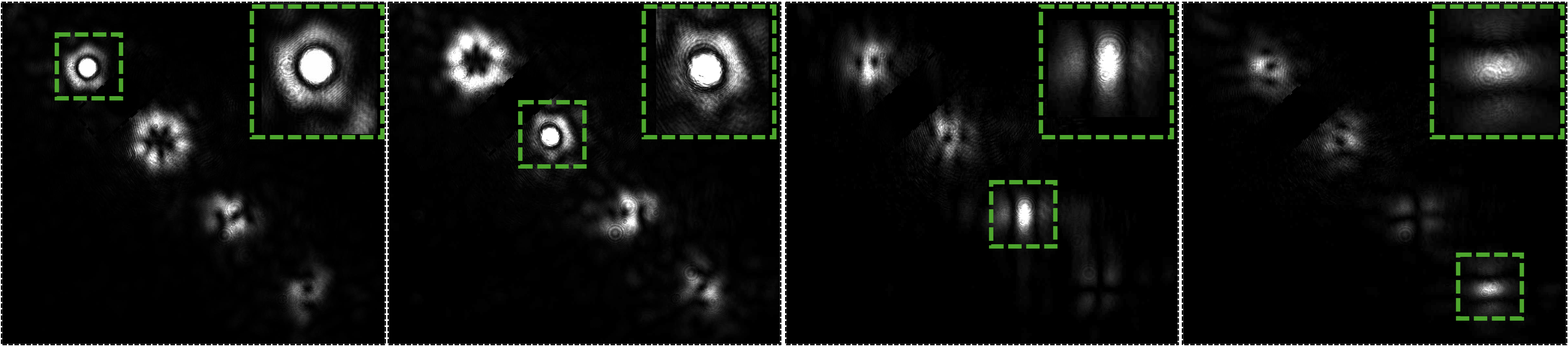}
    \caption{Experimental results for sorting four orthogonal modes, with mixed type: LG modes and two HG: $LG_{0,2}$, $LG_{0,-2}$, $HG_{1,0}$ and $HG_{0,1}$. }
    \label{fig:sortMixLGHG}
\end{figure}

\begin{figure}
    \centering
    \includegraphics[width=1.0\linewidth]{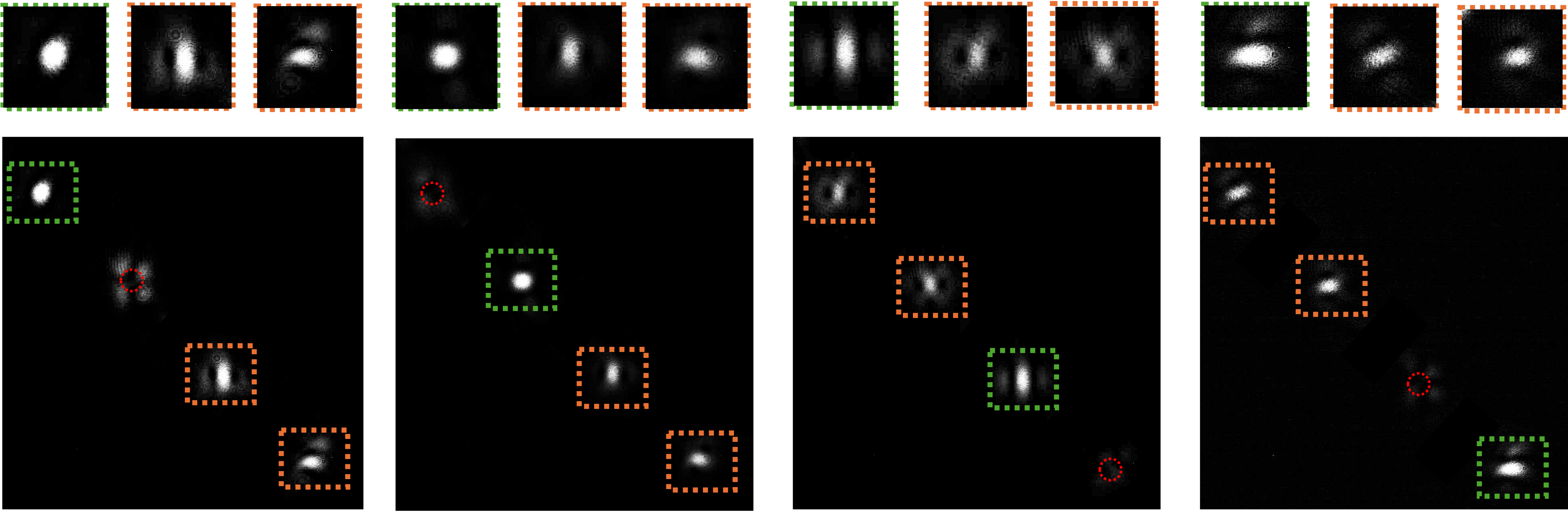}
    \caption{Experimental results for sorting four non-orthogonal modes (MUB): LG modes and two HG: $LG_{0,1}$, $LG_{0,-1}$, $HG_{1,0}$ and $HG_{0,1}$.}
    \label{fig:sortMixMUB}
\end{figure}

As it was explained, the proposed sorter method can possibly support any set of spatial modes. In particular, we could examine the case of a combination between LG and HG modes. Figure \ref{fig:sortMixLGHG}
shows an experimental result sorting two LG and two HG modes, which are orthogonal. In contrary, Figure \ref{fig:sortMixMUB} shows a case where two non-orthogonal modes are sorted.


\newpage

\section{Fork-Shaped Sorter}
\label{app:fork}
As discussed in the paper, our proposed sorter naturally gives rise to a fork-shaped sorter when orbital angular momentum is involved in spatial modes, as shown in Figure~\ref{app:fork}.

\begin{figure}
    \centering
    \includegraphics[width=1.0\linewidth]{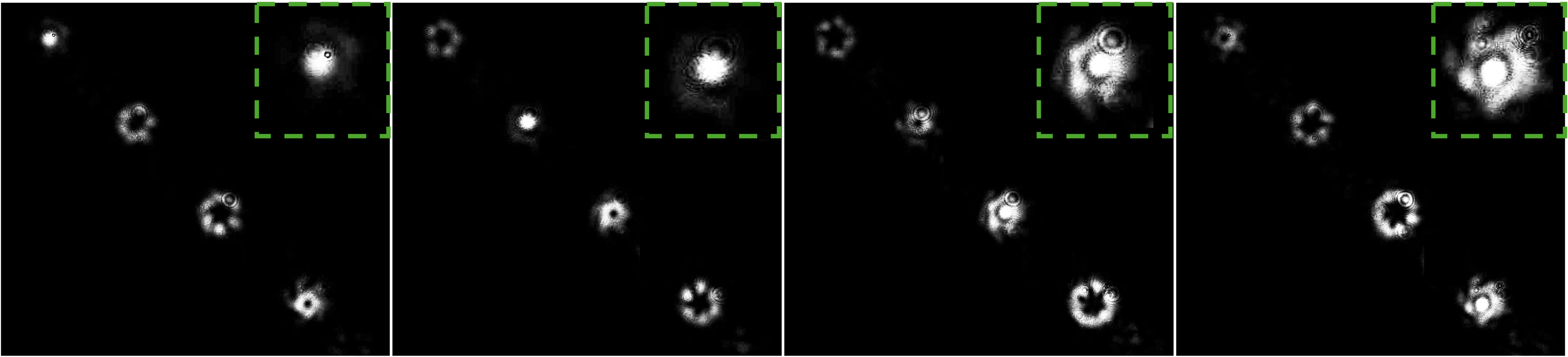}
    \caption{Experimental results of sorting four Laguerre–Gaussian modes with different angular momentum: $LG_{0,1}$, $LG_{0,-1}$, $LG_{0,2}$, and $LG_{0,-2}$.}
    \label{fig:fork_experiment}
\end{figure}

Figure~\ref{fig:fork_experiment} presents the experimental results using our fork-like sorter.

\begin{figure}
    \centering
    \includegraphics[width=1.0\linewidth]{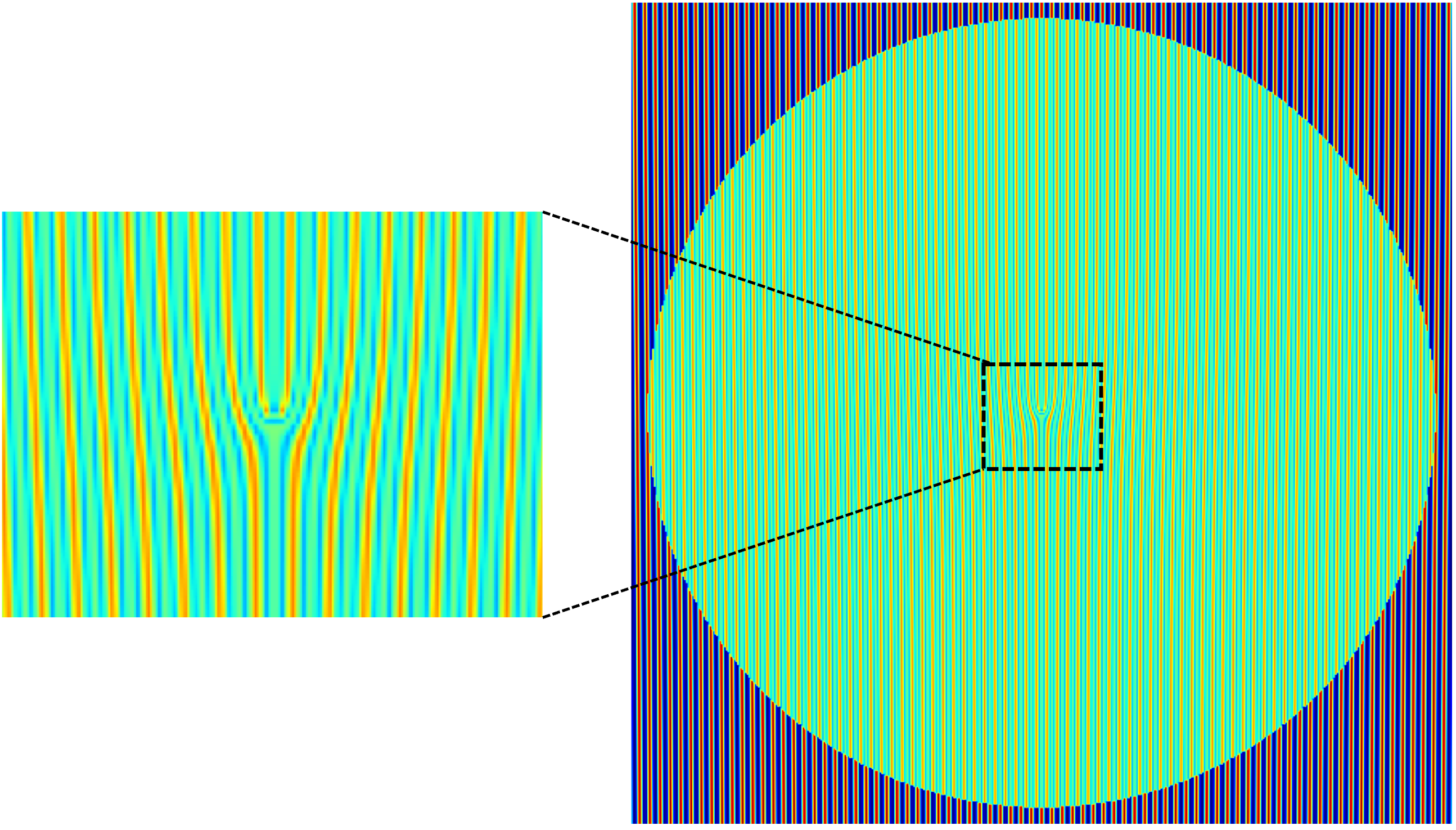}
    \caption{Phase mask of a sorter for four Laguerre–Gaussian beams with different angular orbital momentum (for convenience, the detectors were positioned along a horizontal line). In this particular case, our method gives the well-known Fork sorter and hence perform as a generalization}
    \label{fig:fork}
\end{figure}

\newpage

\section{Wavelength and Noise Analysis}
\label{app:wg_noise_analysis}

\begin{figure}
    \centering
    \includegraphics[width=1.0\linewidth]{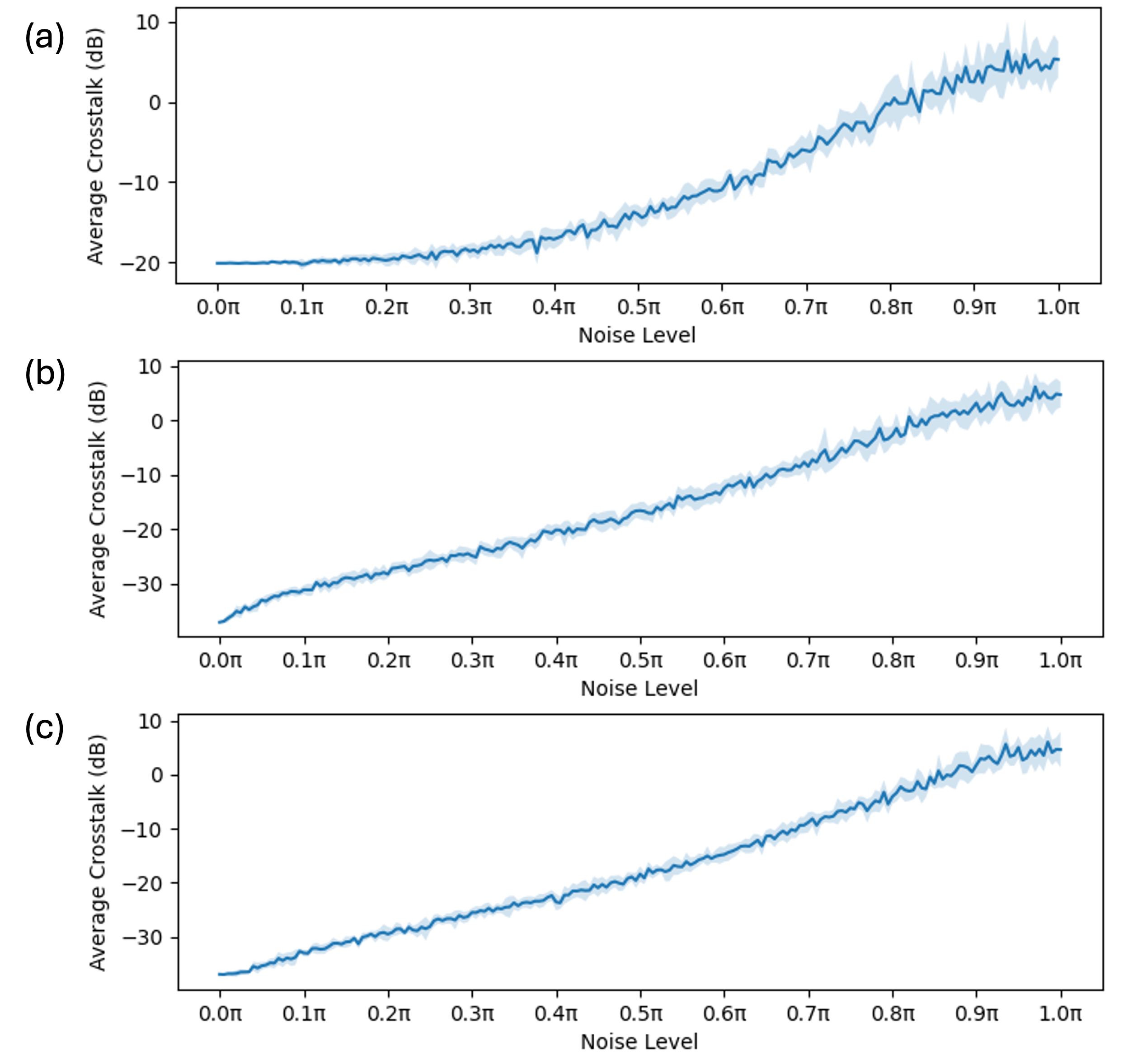}
    \caption{Noise analysis. Average cross-talk versus noise standard deviation for each one of the mode types: (a) Gaussian modes $HG_{0,0},HG_{1,0},HG_{0,1}, HG_{1,1}$ ; (b) Laguerre–Gaussian modes $LG_{0,0},LG_{1,0},LG_{2,0},LG_{3,0}$; and (c) Bessel-Gaussian modes $BG_{0,0},BG_{1,1},BG_{2,-2},BG_{3,3}$.
    The noise model consists of independent Gaussian noise with zero mean and varying standard deviation (shown on the x-axis). The results are averaged over an ensemble of 100 samples, and the shaded area represents the standard deviation. }
    \label{fig:noiseAnalysis}
\end{figure}

Figure \ref{fig:noiseAnalysis} illustrates the system's cross-talk sensitivity to varying levels of random phase noise, modeled as uncorrelated Gaussian noise applied independently to each pixel in a $512 \times 512$ pixel sorter. The results are averaged over an ensemble of 100 random noise realizations, with the shaded regions representing the standard deviation.

\begin{figure}
    \centering
    \includegraphics[width=1.0\linewidth]{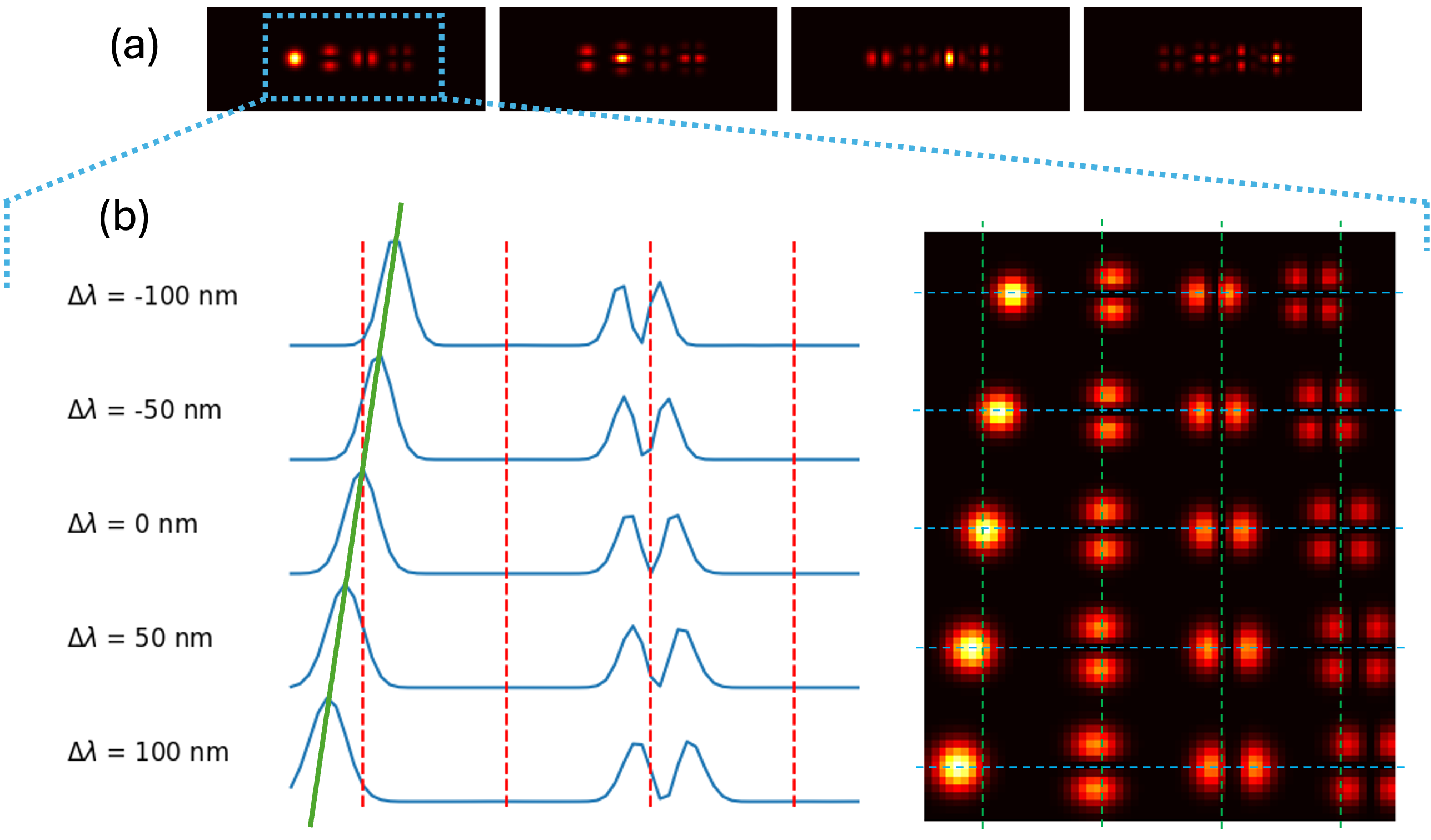}
    \caption{Wavelength analysis. Panel (a) illustrates the projected mask corresponding to each input mode for the Hermite-Gaussian modes $HG_{0,0},HG_{0,1},HG_{1,0}, HG_{1,1}$. In panel (b), we focus solely on the $HG_{0,0}$ mode, examining the projected intensity variations across different wavelengths. The green line illustrate the correction needed to tune the new detector location, for different wavelength. This analysis reveals that a wavelength discrepancy of approximately $50$ to $100$ nm can result in detection errors.}
    \label{fig:wgAnalysis}
\end{figure}

Wavelength sensitivity is evaluated in Figure \ref{fig:wgAnalysis}, which illustrates an example of Hermite-Gaussian mode sorting under wavelength deviations of $\pm 50$ nm and $\pm 100$ nm around the target wavelength of 633 nm.

\subsection{Spectroscopy application}
\begin{figure}
    \centering
    \includegraphics[width=1.0\linewidth]{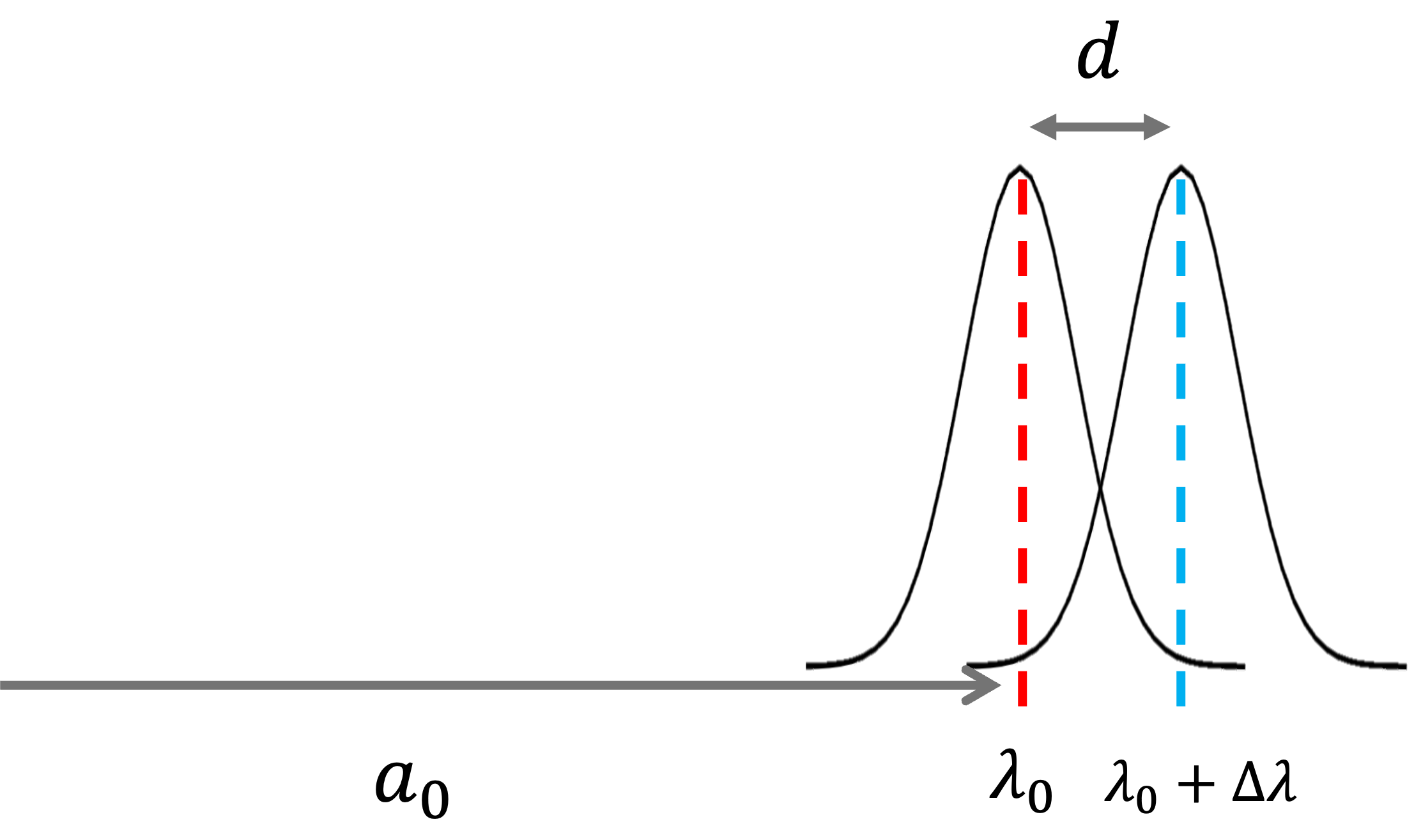}
    \caption{Rayleigh-like separation criteria for spectroscopy. Assuming some mode was sorted for a detector location at a distance $a_0$ from the optical axis. Sorting the same mode, with the same sorter, with a wavelength $\lambda$ gives a shifted delta-function in distance $d$.}
    \label{fig:RayleighLikeSpectroscopy}
\end{figure}

The high spatial precision of the detector's location can be leveraged for spectroscopic applications (see Figure~\ref{fig:RayleighLikeSpectroscopy}). By taking advantage of the scaling effect induced by wavelength variations, we observe that a shift in wavelength results in a corresponding displacement of the spatial sorting location. This relationship can be expressed as:
\begin{equation}
    \left(1 + \frac{\Delta \lambda}{\lambda_0}\right)\left(1 + \frac{d}{a_0}\right) = 1 \ .
\end{equation}
Here, $\lambda_0$ denotes the nominal wavelength for which the sorter was designed, $\Delta \lambda$ represents the deviation from this base wavelength, $a_0$ is the sorting distance from the optical axis at $\lambda_0$, and $d$ is the resulting spatial shift due to the wavelength change. Note that both $\Delta \lambda$ and $d$ may take positive or negative values.

Rearranging the expression yields:
\begin{equation}
    d = a_0 \left( \frac{1}{1 + \tfrac{\Delta \lambda}{\lambda_0}} - 1 \right) \ .
\end{equation}

Assuming the detector has a pixel size of $p$, the minimum detectable wavelength shift corresponds to a spatial displacement satisfying:
\begin{equation}
    p \leq a_0 \left| \frac{1}{1 + \tfrac{\Delta \lambda}{\lambda_0}} - 1 \right| \ .
\end{equation}

As a concrete example, for $a_0 = 1~\mathrm{mm}$, $\Delta \lambda = 1~\mathrm{nm}$, and $\lambda_0 = 600~\mathrm{nm}$, the required pixel resolution is:
\begin{equation}
    p \leq 1.6~\mu\mathrm{m} \ .
\end{equation}
Another way to formulate this:
\begin{equation} \label{eq:spectroscopy}
    \Delta \lambda  = \lambda_0 \left( \frac{1}{1+\tfrac{d}{a_0} }-1 \right) \ .
\end{equation}
Or, for minimum separation of $p=|d|$:
\begin{equation}
    \lambda_0 \left( \frac{1}{1+\tfrac{p}{a_0} }-1 \right) \leq \Delta \lambda  \leq \lambda_0 \left( \frac{1}{1-\tfrac{p}{a_0} }-1 \right) \ .
\end{equation}
For example, given pixel size of $p=10_{\mu m}$, and $\lambda_0 = 633_{nm}$ , and $a_0 = 1_{mm}$:
\begin{equation}
    -6.27 _{nm} \leq \Delta \lambda  \leq 6.39_{nm} \ .
\end{equation}

To assess our proposed spectroscopy technique, we utilized the sorter for Bessel-Gaussian modes, specifically $BG_{0,0}$, $BG_{1,1}$, $BG_{2,-2}$, and $BG_{3,3}$, concentrating on the sorting location of $BG_{1,1}$. Each time, the sorter was fine-tuned for another individual wavelengths and consistently employed a 633$_{nm}$ laser for each experiment. For each sorting configuration, we determined the distance from the optical axis, denoted as $a_{\text{pixels}}$, and calculated the presumed $\Delta \hat{\lambda}$ (using eq. \ref{eq:spectroscopy}) from this data. The outcomes are presented in table \ref{tab:spectroscopy}.
\begin{table}
    \centering
    \begin{tabular}{|c|c|c|c|c|l|}\hline
         $\lambda_{ \text{nm} }$&  633&  640&  650& 660&670\\\hline
 $\Delta \lambda _{ \text{nm} }$& \textbf{0}& \textbf{7}& \textbf{17}& \textbf{27}& \textbf{37}\\\hline
         $a_{\text{pixels}}$&  811.089&  804.006&  789.143&  777.165&765.813\\\hline
         $d/a_0$&  1&  -0.008733&  -0.027057&  -0.041825&-0.055821\\\hline
 $\Delta\hat{ \lambda }_{ \text{nm} }$& \textbf{0} & \textbf{5.577} & \textbf{17.604} & \textbf{27.631} & \textbf{37.424}\\\hline
 Error ($\%$)& 0& 20& 3.5& 2.3&1.1\\ \hline
    \end{tabular}
    \caption{Spectroscopy results for sorting Bessel-Gaussian modes ($BG_{0,0}$, $BG_{1,1}$, $BG_{2,-2}$, and $BG_{3,3}$), with a focus on the mode $BG_{1,1}$. The column $a_{\text{pixels}}$ indicates the measured distance (in pixels) from the optical axis, while $\Delta \hat{\lambda}$ represents the estimated wavelength shift inferred from the pixel displacement. The relative error is calculated as $\tfrac{|\Delta \lambda - \Delta \hat{\lambda}|}{|\Delta \lambda|}$.}
    \label{tab:spectroscopy}
\end{table}

\end{document}